\documentclass[prc,floatfix,nofootinbib,showpacs,showkeys,preprint]{revtex4-1}

\usepackage{url}
\usepackage{pdfsync}
\usepackage{slashed}
\usepackage{graphicx}%\usepackage{subcaption}

\begin{document}

% Page header
\markboth{U. Mosel}{Neutrino Interactions}

% Title
\title{Neutrino Interactions with Nucleons and Nuclei:\\ Importance for Long-Baseline Experiments}

%Authors, affiliations address.
\author{Ulrich Mosel}
\affiliation{Institut fuer Theoretische Physik, Universitaet Giessen, Germany, D-35392}
\email[Contact e-mail: ]{mosel@physik.uni-giessen.de}

%Abstract
\begin{abstract}
This article reviews our present knowledge of neutrino interactions with nucleons and discusses the interactions with nuclei, the target material of all presently running and planned long-baseline experiments. I emphasize descriptions of semi-inclusive reactions and full descriptions of the final state; the latter are needed to reconstruct the incoming neutrino energy from final-state observations. I then discuss Monte Carlo generator and more advanced transport theoretical approaches in connection with experimental results on various reaction mechanisms. Finally, I describe the effects of uncertainties in the reconstruction of the incoming neutrino energy on oscillation parameters. The review argues that the precision era of neutrino physics also needs precision-era generators.
\end{abstract}

%Keywords, etc.
%\begin{keywords}
%neutrino, neutrino interaction, long-baseline experiments
%\end{keywords}
%\maketitle

%Table of Contents

\maketitle
\tableofcontents

% Heading 1
\section{INTRODUCTION}
The interactions of neutrinos with \emph{nucleons} can provide valuable information about axial properties and transition form factors.  For example, the nucleon's axial form factor is still poorly known. It is usually reduced to a dipole ansatz, with one free parameter, the axial mass. This axial mass ($M_A$) has been determined in many neutrino experiments on nucleons (or deuterons) and assumes a value of approximately 1 GeV \cite{Bernard:2001rs}. The assumed dipole form of the axial vector form factors, however, cannot be checked further by experiment; the vector form factors obtained from electron scattering show a significantly more complicated dependence on the squared four-momentum transfer $Q^2$ \cite{Arrington:2006zm}. The transition form factors to nucleon resonances are even less known. For example, for the $\Delta$ resonance the transition current involves three vector form factors and three axial ones. Whereas the three vector form factors are reasonably well determined by electron-induced pion production on the nucleon, the three axial form factors are largely unknown. Present data seem to be sensitive to only one of them as discussed below in Sect.\ \ref{s:pionprod}.

The investigation of interactions of neutrinos with \emph{nuclei} is interesting from the point of view of nuclear many-body theory (NMBT). It can provide valuable information on the electroweak response of nuclei to axial perturbations and, thus, supplement our previous knowledge from electron scattering experiments. It is also interesting from a practical point of view, with regard to long-baseline experiments, such as T2K, MINOS, NOvA and the future DUNE (formerly called LBNE), that attempt to extract neutrino properties from the observation of neutrino oscillations. In these experiments the event rate (flux multiplied by cross section) at a given neutrino energy $E_\nu$ at a far detector is compared with that at a near detector at the same energy. From that comparison one can extract the neutrino oscillation parameters, mixing angles, and possibly a CP-invariance violating phase. The complication lies in the fact that the neutrino energy is not known because of the special production method of neutrinos as secondary decay products of hadrons, mostly pions and kaons, that were produced in primary reactions of protons with nuclei. The neutrino energy thus must be reconstructed event by event from the final state of the reaction, at both the near and the far detectors.

Because all modern experiments use nuclear targets, such as H$_2$O, CH$_{n}$ and $^{40}$Ar,  the energy reconstruction depends not only on the initial neutrino-nucleus interaction but also on the final-state interactions (FSI) of all particles. The precision with which neutrino oscillation properties can be extracted from such experiments then depends directly on the description of the final state of the neutrino-nucleus interaction.

\begin{figure}[!ht]
\begin{center}
\includegraphics[width=0.8\columnwidth]{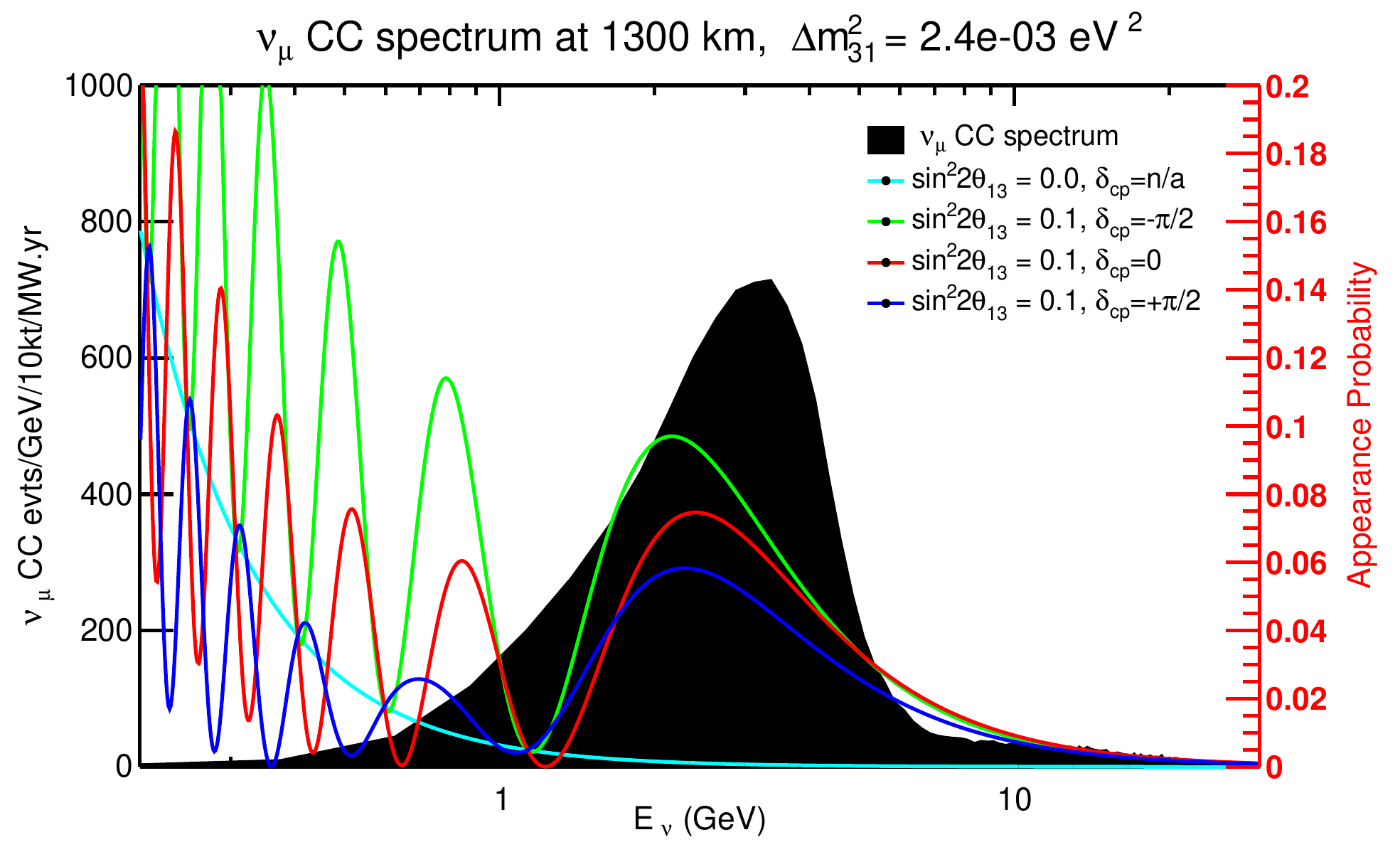}
\caption{Appearance probability of $\nu_e$ in a $\nu_\mu$ beam at a distance of 1300 km, calculated for standard oscillation mixing angles. The four colored curves illustrate the sensitivity of the expected signal to the neutrino mixing angle $\theta_{13}$ and the CP-violating phase $\delta_{CP}$. The black peak shows the expected energy distribution for the $\mu$-neutrino beam. From Reference \cite{Adams:2013qkq}.}
\label{fig:LBNE-oscill}
\end{center}
\end{figure}

%\begin{textbox}[h]\section{NEUTRINO BEAM ENERGY}
%Unlike in any other nuclear physics experiment in neutrino-induced reactions the beam energy is not known but must be reconstructed from the final state of the %reaction. The accuracy of that reconstruction affects the extraction of neutrino oscillation parameters.
%\end{textbox}
To get a sense for the accuracy needed for the energy reconstruction in oscillation experiments, it is helpful to look at {\bf Fig. \ref{fig:LBNE-oscill}}. The figure shows the expected oscillation signal for DUNE as as a function of neutrino energy $E_\nu$ for some values of two neutrino properties: the mixing angle $\theta_{13}$ and the CP-violating phase $\delta_{CP}$.
The three curves under the flux profile can be distinguished from one another only if the neutrino energy can be determined to better than approximately 100 MeV,  which provides a first hint to the accuracy needed for the energy reconstruction at DUNE.
%\begin{marginnote}[120pt]
%\entry{DUNE}{Deep Underground Neutrino Experiment}
%\end{marginnote}

The focus of this review is on our understanding of neutrino-nucleon and neutrino-nucleus interactions and their effect on the neutrino energy reconstruction. After a brief review of neutrino interactions with nucleons, I provide a short overview of the theory needed to describe interactions with nuclei and describe the important role of FSI. I then discussion the importance of these interactions for the energy reconstruction in long-baseline experiments. I also discuss the need for better nuclear theory-based generators necessary to achieve higher precision in the extraction of neutrino parameters.

For a summary of experimental results on neutrino interactions I refer to two recent review articles \cite{Gallagher:2011zza}, \cite{Formaggio:2013kya}; the former focusses on quasi-elastic (QE) scattering cross sections whereas the latter provides a rather complete summary of experimental cross sections in various energy regimes. Another review \cite{Conrad:1997ne} gives an excellent presentation of neutrino interactions at high energies. Another classical resource is an extended review article by LLewellyn Smith \cite{LlewellynSmith:1971zm}, which contains theoretical and experimental developments that are still relevant today. A more modern review of neutrino interactions, mainly from a theoretical point of view, is that by Alvarez-Ruso et al.\ \cite{Alvarez-Ruso:2014bla}. Finally, another review in this volume \cite{Rubbia:2016} covers modern long-baseline experiments.

%Heading 1
\section{INTERACTIONS\ WITH\ NUCLEONS}
In this section I discuss our present understanding of reactions on the nucleon for the three major reaction processes.
% Heading 2
\subsection{Quasi-Elastic Scattering}
QE Scattering on an isolated nucleon, as for a charged-current interaction of an incoming antineutrino of flavor $l$ on a proton with an outgoing neutron and a lepton $l$
\begin{equation}     \label{trueQE}
\bar\nu_l + p \rightarrow n + l
\end{equation}
is a very simple reaction. The corresponding reaction of a neutrino on a neutron is already more complicated because there are no neutron targets. The use of a deuteron target instead already requires some nuclear structure information, in this case about the energy and momentum distribution of the neutron in the deuteron and about the reaction mechanism. Due to the two-body kinematics in Eq.\ \ref{trueQE} the measurement of the energy and angle of the outgoing lepton $l$ also determines the incoming energy and the momentum transfer. Essential for this result is that the final state can unequivocally be identified such that only one nucleon and a lepton (and, for example, no pion) are present.

The cross section for the QE scattering (\ref{trueQE}) is obtained by contracting the hadron current with the lepton current\footnote{A summary of all essential theoretical formulas can be found in References \cite{Leitner:2009zz,Leitner:2006ww,Leitner:2006sp}}. The relevant vertex function $\Gamma$ is given by a combination of a vector current ($V$) and an axial current ($A$)
\begin{equation}
\Gamma^\mu_{\rm QE} = V^\mu_{\rm QE} - A^\mu_{\rm QE} ~,
\end{equation}
with
\begin{eqnarray}
V^\mu_{\rm QE} &=& F_1 \gamma^\mu + \frac{F_2}{2M} {\rm i} \sigma^{\mu \lambda} q_\lambda \\
A^\mu_{\rm QE} &=&  - F_{\rm A} \gamma^\mu\gamma^5 - \frac{F_{\rm P}}{M} q^\mu \gamma^5 ~.
\end{eqnarray}
Here $M$ is the nucleon mass and all the vector form factors $F_{1,2}$ and the axial form factors $F_{\rm A}$ and $F_{\rm P}$ depend on the square of the four-momentum transfer $Q^2= -q^2$ alone, because for QE scattering the energy transfer is fixed to $\omega = Q^2/(2M)$. The vector form factors are directly related to the electromagnetic form factors and can be determined by electron scattering; a recent fit can be found in References \cite{Arrington:2006zm,Bodek:2010km,Pacetti:2015iqa}. The form factor $F_{\rm P}$ can be related to $F_{\rm A}$ by invoking pion pole dominance so that only one axial form factor $F_{\rm A}(Q^2)$ is needed.

Theory alone says little about the detailed shape of the axial form factor $F_{\rm A}$. The vector meson dominance hypothesis predicts that it should be the sum of many monopoles with isovector axial vector masses whereas perturbative QCD (pQCD) arguments predict that asymptotically, for large $Q^2$, the form factor should go as $1/Q^4$ \cite{Masjuan:2012sk}. This asymptotic behavior could  appear if close-lying monopoles, such as those corresponding to the two lowest axial vector mesons, conspire in their coupling such that the result is a dipole\footnote{The classical analogue is that of an electrostatic dipole potential which emerges when two charges of equal strength, but opposite sign, are positioned close to each other.}
\begin{equation}   \label{dipole}
F_{\rm A}(Q^2) = \sum_i \frac{\alpha_i}{1 + \frac{Q^2}{m_i^2}} \longrightarrow F_{\rm A}(Q^2) = \frac{g_{\rm A}}{(1 + \frac{Q^2}{M_{\rm A}})^2}.
\end{equation}
In this case the axial form factor's shape is determined by only one parameter, the axial mass $M_{\rm A}$. While Eq.\ \ref{dipole} indeed fulfills the asymptotic requirement, in the experimentally relevant region of low $Q^2$ the form factor could have a different shape \cite{Gari:1984qs,Bhattacharya:2011ah,Bhattacharya:2015mpa,Amaro:2015lga}.

\begin{figure}[h]
\includegraphics[width=3in]{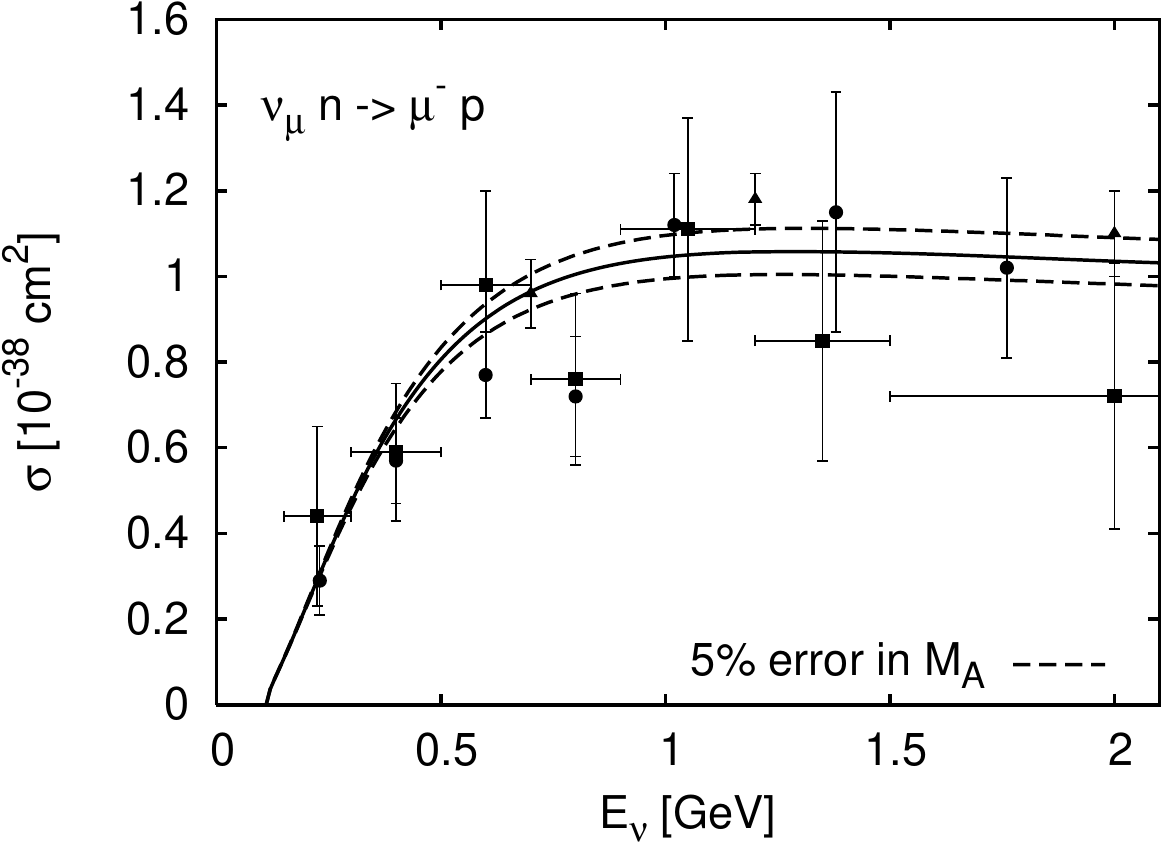}
\caption{Charged-current quasi-elastic cross section for $\nu_\mu$ scattering off neutrons. The experimental error bars are clearly much larger than the uncertainties due to using different values for $M_{\rm A}$; the large error bars also lead to a correspondingly large uncertainty in the shape. Data are from References \cite{Barish:1977qk,Mann:1973pr,Baker:1981su}. From Reference\cite{Leitner:2006ww}}
\label{fig:QE}
\end{figure}

Nearly all analyses of neutrino QE data have used the dipole form factor. The axial mass extracted usually shows large error bars because all the experiments with elementary targets ($p, D$) were done approximately 35 years ago with relatively weak neutrino currents. The world average for the extracted axial mass is 1.03 GeV; the value extracted from charged pion electroproduction experiments, which are also sensitive to $F_{\rm A}$, is close \cite{Bernard:2001rs}. \textbf{Figure \ref{fig:QE}} illustrates the sensitivity of the total QE cross section as a function of neutrino energy to the axial mass.

\subsection{Pion Production}   \label{s:pionprod}
At energies above approximately 200 MeV the first inelastic excitations of the nucleon connected with pion production become possible. Most of the nucleon resonances have spin $1/2$ and $3/2$. The transition currents to the spin-1/2 resonances have the same form as given above for the nucleon. The hadronic transition currents to the 3/2-resonances, by contrast, have a much more complicated structure. Among these at the lower energies pion production through the $\Delta(1232)$ resonance with spin 3/2 and isospin 3/2 ($J,T = 3/2,3/2$) is dominant.

The two currents are given by \cite{Albright:1965zz}
\begin{widetext}
\begin{eqnarray}    \label{Deltacurr}
V^{\alpha \mu}_{3/2} &=& \frac{C_3^V}{M} \left( g^{\alpha \mu} \slashed{q} - q^\alpha\gamma^\mu \right) + \frac{C_4^V}{M^2} \left( g^\alpha\mu q \cdot p´ - q^\alpha p´\mu\right) + \frac{C_5^V)}{M^2}\left( g^{\alpha \mu} q \cdot p - q^\alpha p^\mu \right) + g^{\alpha \mu} C_6^V \nonumber \\
A^{\alpha \mu}_{3/2} &=& - \left[ \frac{C_3^A}{M} \left( g^{\alpha \mu} \slashed{q} - q^\alpha\gamma^\mu \right) + \frac{C_4^A}{M^2} \left( g^{\alpha\mu} q \cdot p´ - q^\alpha p^\mu \right) + C_5^A g^{\alpha \mu} + \frac{C_6^A}{M^2} q^\alpha q^\mu \right] \gamma^5 ~.
\end{eqnarray}
\end{widetext}
They enter via the vertex factor $\Gamma$ for a positive-parity $J = 3/2$ resonance
\begin{equation}
\Gamma^{\alpha \mu} = \left( V^{\alpha \mu} - A^{\alpha \mu} \right) \gamma^5
\end{equation}
into the hadronic tensor, which is given by
\begin{equation}
H^{\mu \nu} = \frac{1}{2} {\rm Tr} \left[\slashed{p} + M) \Gamma^{\alpha \mu} \Lambda_{\alpha \beta} \Gamma^{\beta \nu}\right] ~,
\end{equation}
where $\Lambda_{\alpha \beta}$ is (for the $\Delta$) the spin-3/2 projector.
Contracting this with the lepton tensor gives, as usual, the resonance production cross section
\begin{equation}       \label{resprod}
\frac{{\rm d}\sigma^{\rm med}}{{\rm d}\omega {\rm d}\Omega'} = \frac{|\mathbf{k}'|}{32 \pi^2} \frac{\mathcal{P}^{\rm med}(p')}{[(k \cdot p)^2 - m_\ell^2M^2]^{1/2}} \left| \mathcal{M}_R \right|^2 ~.
\end{equation}
Here $p$ denotes the nucleon's four-momentum, $p'$ that of the outgoing resonance and $k$ and $k'$ are those of the initial and final state lepton, respectively. The quantities $\omega$ and $\Omega'$ represent the energy transfer and the scattering angle of the outgoing lepton, respectively. The spectral function (SF) of the resonance is denoted by $\mathcal{P}(p')$; in medium it can differ from the free SF and, therefore, carries the supersript 'med' in Eq.\ \ref{resprod}. To obtain the pion production cross section one multiplies the resonance formation cross section with the branching ratio for decay into the $\pi N$ channel
\begin{equation}    \label{sigma-pi}
\frac{{\rm d}\sigma^{\rm med}}{{\rm d}\omega {\rm d}\Omega' {\rm d}\Omega_\pi^{CM}} =  \frac{1}{4 \pi} \frac{{\rm d}\sigma^{\rm med}}{{\rm d}\omega {\rm d}\Omega'} \frac{\Gamma_{R \to N\pi}}{\Gamma_{\rm tot}} ~.
\end{equation}

The vector form factors $C_i^V(Q^2)$ in Eq.\ \ref{Deltacurr} are directly related to the electromagnetic transition form factors \cite{Leitner:2009zz}. They can be obtained from the measured helicity amplitudes, determined in, e.g., the MAID analysis \cite{Drechsel:2007if}. The often-used Rein-Sehgal model for the form factors \cite{Rein:1980wg} is known to fail in its description of electron scattering data \cite{Graczyk:2007bc,Leitner:2008fg}.

\begin{figure}[!hbt]
\begin{minipage}[c]{0.5\textwidth}
\includegraphics[width=\textwidth]{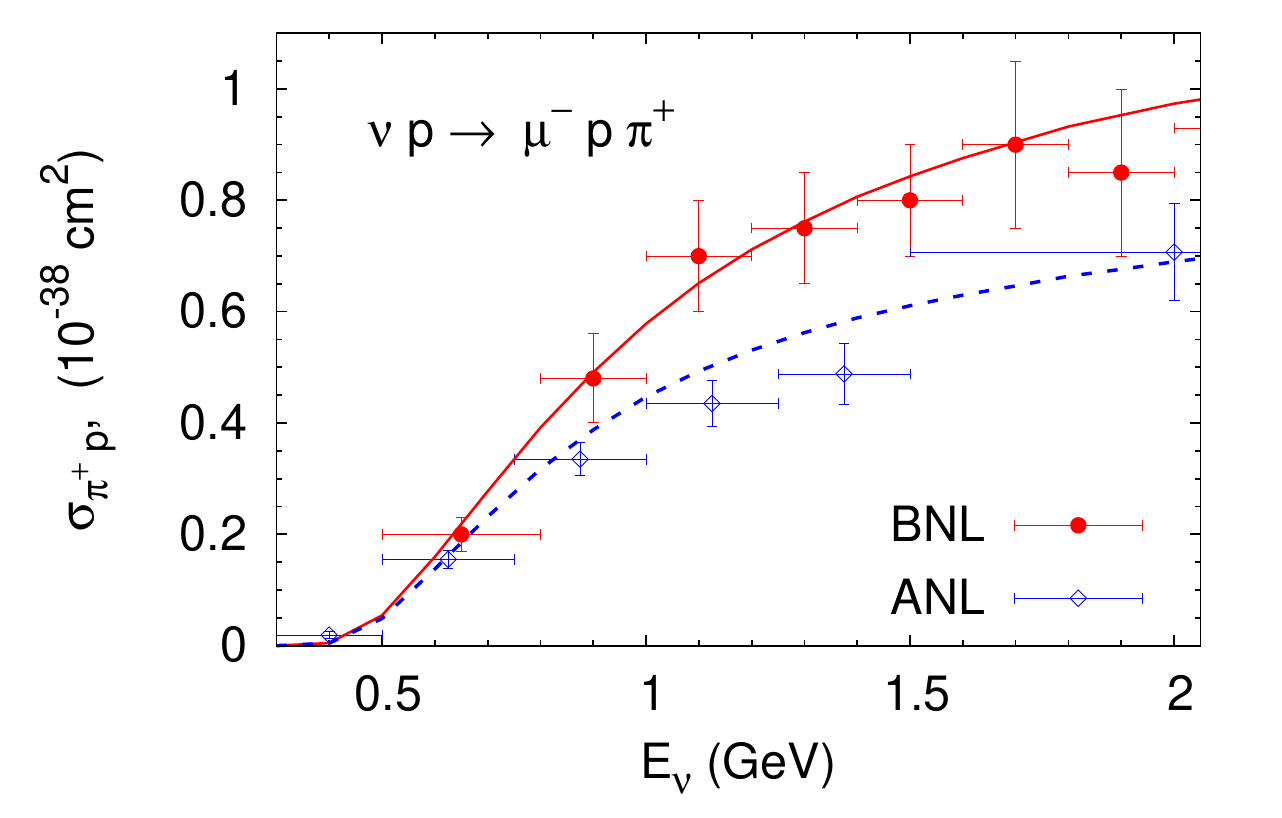}
\end{minipage}
\hfill
\begin{minipage}[c]{0.46\textwidth}
\includegraphics[width=\textwidth]{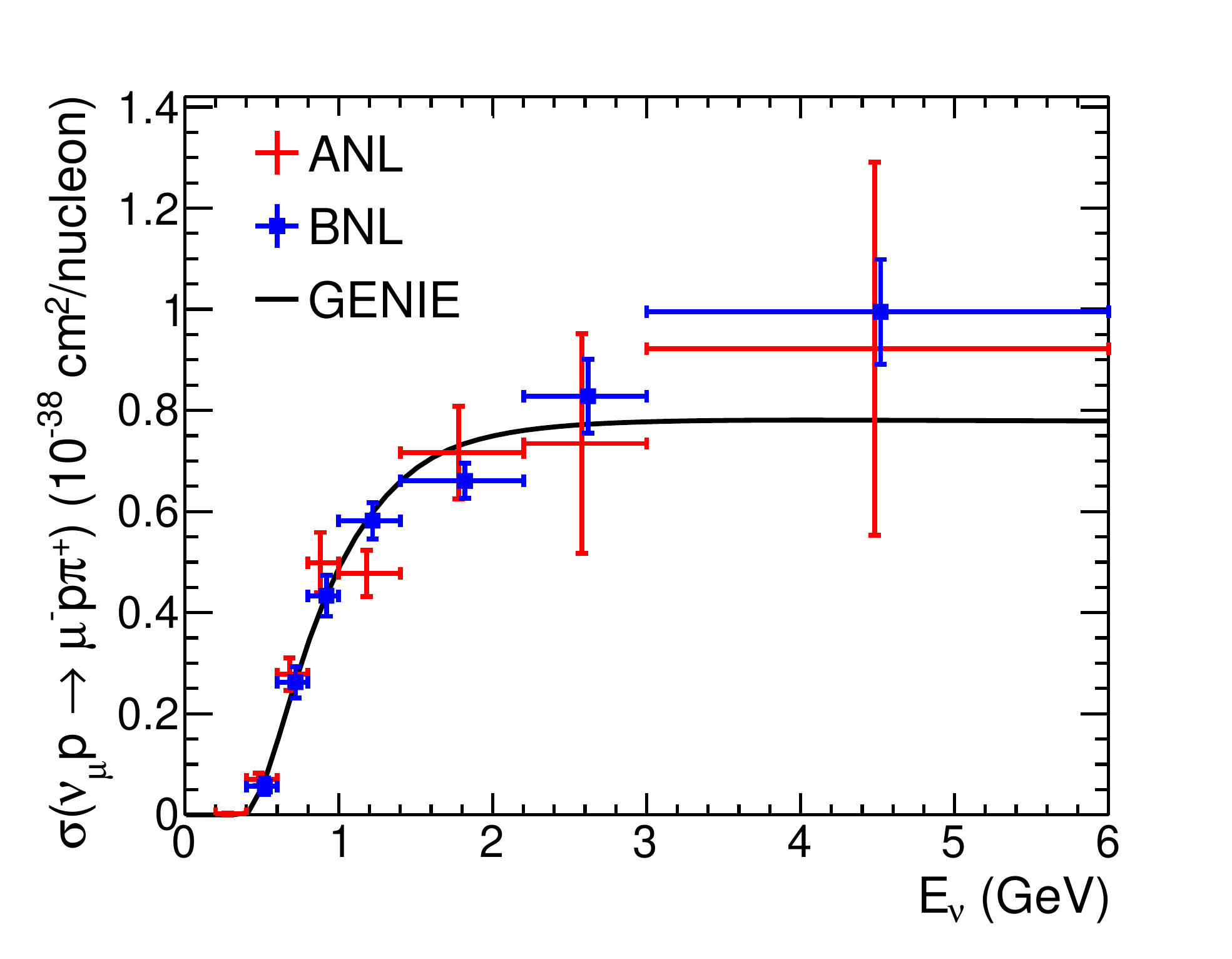}
\end{minipage}
\caption{Left: Single-pion production cross sections on a proton target obtained in the BNL \cite{Kitagaki:1986ct} (circles; solid curve) and the ANL experiments \cite{Radecky:1981fn}  (diamonds; dashed curve). The curves give the lower (ANL-tuned) and upper (BNL-tuned) boundaries on the elementary input as used in GiBUU (from \cite{Lalakulich:2012cj}). Right: same cross sections after flux recalibration of the BNL cross sections (from \cite{Wilkinson:2014yfa}. The solid curves represent models implemented in neutrino generators.}
\label{fig:pi}
\end{figure}

The only data sets available for pion production on a nucleon are those obtained at Argonne National Laboratory (ANL) \cite{Radecky:1981fn} and Brookhaven National Laboratory (BNL) \cite{Kitagaki:1986ct}.  \textbf{Figure \ref{fig:pi}}(left) shows
that these data sets differ by approximately 30\% at the higher energies. The quality of these data is obviously not sufficient to determine all four axial form factors $C_i^A(Q^2)$.

Already in 1965 researchers had noticed that $C_5^A$ gives the dominant contribution \cite{{Albright:1965zz}}. $C_6^A$ can be related to $C_5^A$ by partial conservation of the axial currenct (PCAC) \cite{Lalakulich:2005cs} and $C_3^A$ is set to zero according to an old analysis by Adler \cite{Adler:1968tw}, whereas $C_4^A$ is linked to $C_5^A$. On the basis of these relations all theoretical analyses have used only one axial form factor $C^A_5(Q^2)$, with various parameterizations that usually go beyond that of a simple dipole $C_5^A(Q^2)$ \cite{Leitner:2006ww,Leitner:2006sp,Lalakulich:2005cs,Graczyk:2009qm,Hernandez:2010bx}.

Both the ANL and BNL experiments extracted various invariant mass distributions from their data. The analysis of these invariant mass data together with the experimental $d\sigma/dQ^2$ distributions led Lalakulich et al.\ \cite{Lalakulich:2010ss} to conclude that the BNL data were probably too high. This conclusion has been confirmed by a reanalysis of the old data by Wilkinson et al \cite{Wilkinson:2014yfa} who used the QE data obtained in the same experiment for a flux calibration. After that flux recalibration the BNL data agreed with the ANL data (see right part of \textbf{Figure \ref{fig:pi}}).

A complication in determining the resonance parameters is the presence of background amplitudes which have been explored in effective field theory models \cite{Hernandez:2007qq,Lalakulich:2010ss,Alvarez-Ruso:2015eva}. Significantly more complicated is the dynamical coupled-channel model of photopion, electropion and weak pion production \cite{Nakamura:2015rta} that has been applied to all resonances with invariant masses up to 2.1 GeV. In this model background and resonance contributions emerge from the same Lagrangian. It is puzzling that these calculations give a cross section that is close to the higher-lying BNL cross sections for single pion production. New measurements using elementary targets are thus needed to solve this problem.

\subsection{Deep-Inelastic Scattering}
\begin{figure}[h]
\includegraphics[width=3.5in]{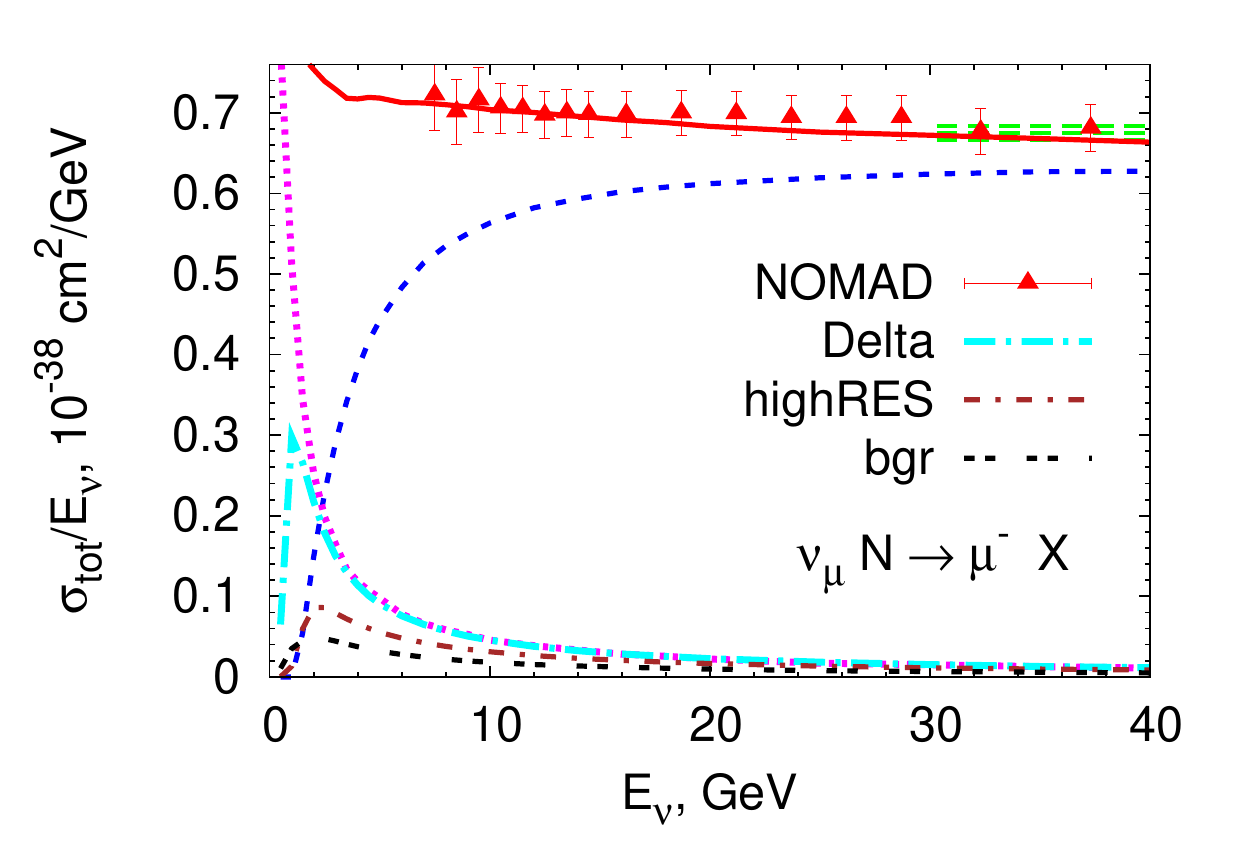}
\caption{Charged-current neutrino interaction with an isoscalar target. Data are from the NOMAD experiment \cite{:2007rv}, the curves stem from a Giessen-Boltzmann-Uehling-Uhlenbeck calculation \cite{Lalakulich:2012gm}. The various channels that contribute to the cross section are given as follows: The pink dotted curve represents the contribution from quasi-elastic scattering and the blue, dashed curve represents that from deep-inelastic scattering. 'Delta' denotes excitations of the $\Delta$ resonance, 'HighRes' denotes those from higher lying nucleon resonances and 'bgr' denotes those from background terms in the resonance region. From \cite{Lalakulich:2012gm}.}
\label{fig:DIS}
\end{figure}

Deep-inelastic scattering (DIS) on the nucleon is well defined only in the very high energy regime. Above a neutrino energy of approximately 20 - 30 GeV the cross section is dominated by DIS, namely scattering of the incoming neutrino on individual partons. A detailed discussion of the underlying theory and of many experimental results can be found elsewhere \cite{Conrad:1997ne}. For lower energies of a few GeV many other reaction channels, especially pion production through the $\Delta$ and higher resonances, contribute (\textbf{Figure \ref{fig:DIS}}).
Therefore, the neutrino energy region of approximately 1 - 5 GeV (the so-called shallow-inelastic region) is more difficult to describe as the $Q^2$ range in this regime is wide enough to mix resonance excitations with pQCD processes.
%\begin{marginnote}[220pt]
%	\entry{SIS}{Shallow Inelastic Scattering describes the invariant nucleon mass region %between resonance physics and true perturbative QCD scattering.}
%\end{marginnote}

Usually the inclusive DIS (high-energy) part of the cross section is described by the three structure functions $F_i(Q^2,x)$ \cite{Leader:1996}; here $x = Q^2/(2 M \omega)$. For these excellent empirical fits exist \cite{Bodek:2010km}. The structure functions determine the cross section for inclusive events \cite{Conrad:1997ne}. In order to obtain cross sections for the full event, high-energy event generators such as PYTHIA \cite{Sjostrand:2006za} are used to model the reaction of the incoming gauge boson on partons inside the nucleon. Such generators obtain the final particle yield from a string breaking mechanism. The shallow inelastic region is then treated by some interpolation scheme between a resonance model and a parton model.

\section{INTERACTIONS\ WITH\ NUCLEI}
Interactions of neutrinos with nuclei supplement the information gained by electron scattering off nuclei on nuclear ground- and excited-state properties and reaction mechanisms. They also yield insight into the axial response of nuclei. Moreover, understanding these interactions with nuclei is essential for the reconstruction of the incoming neutrino energy and, thus, for the precision with which oscillation parameters can be extracted from long-baseline experiments. For the latter we need theoretical descriptions of the $\nu A$ interactions that describe the complete final state of the reaction.

A fully quantum-mechanical approach to the problem of describing the final state of a neutrino-nucleus reaction is possible for inclusive and semi-inclusive reactions.  In this case one could use methods from standard reaction theory, such as the Glauber approximation \cite{Martinez:2005xe} or an optical model description \cite{Meucci:2011vd,Meucci:2012yq}. These methods can describe the attenuation in a given channel, but they do not provide any information about where the flux goes.

\subsection{Dynamics of Neutrino--Nucleus Interactions}  \label{subsect:transport}

The only known method of treating the time development of a quantal nuclear many-body system from its initial reaction all the way to the final state with  high particle multiplicity is transport theory \cite{DeGroot:1980dk,Danielewicz:1982kk}. The theoretical basis of this approach was laid more than 50 years ago by Kadanoff \& Baym \cite{Kad-Baym:1962}. Transport theory is widely used in other fields of physics \cite{DeGroot:1980dk}. In nuclear physics it has been used to describe heavy-ion reactions with their inherent large particle multiplicity in the final state \cite{Danielewicz:1982ca,Wang:1991sj,Buss:2011mx,Bertsch:1988ik,Knoll:2011zz}, as well as neutrino transport in supernovae \cite{DeGroot:1980mh,Zhang:2013lka}.
%\begin{textbox}[h]\section{Transport Theory}
%Transport theory is used in such different applications as neutrino transport in dense matter and descriptions of heavy-ion reactions looking for the quark-gluon %plasma.
%\end{textbox}

The Kadanoff--Baym (KB) equations in their gradient expansion form, and using the Botermans--Malfliet off-shell term \cite{Botermans:1990qi}, allow one to propagate the eight-dimensional phase-space distributions $F(x,p)$ of any off-shell particle under the influence of a mean field and interactions with other particles; here $x$ and $p$ are the four-vectors of space-time and momentum-energy, respectively. Essential quantal effects, such as off-shellness of bound hadrons, nuclear binding and the Pauli-principle, are all contained in this method.

%\begin{marginnote}[120pt]
%	\entry{KB Equations}{Kadanoff--Baym equations describe the time development of the Wigner-transform of the nuclear one-body density matrix}
%\end{marginnote}
The KB equations, with the Botermans-Malfliet approximation \cite{Botermans:1990qi}, are given by \cite{Buss:2011mx}
\begin{widetext}
\begin{equation}  \label{KBeq}
\mathcal{D} F(x,p) - {\rm tr}\left\{ \Gamma f,{\rm Re} S^{\rm ret}(x,p)\right\}_{\rm PB} = C(x,p) ~,
\end{equation}
where
\begin{equation}
\mathcal{D} F(x,p) = \left\{ p_0 - H,F\right\}_{\rm PB} = \frac{\partial (p_0 - H)}{\partial x}\frac{\partial F}{\partial p} - \frac{\partial (p_0 - H)}{\partial p}\frac{\partial F}{\partial x}
\end{equation}
\end{widetext}
represents the so-called drift term that determines the propagation under the influence of a mean-field Hamiltonian $H$; the subscript PB denotes a Poisson bracket. In Eq.\ \ref{KBeq} $\Gamma$ is the width of the propagated particle and $S$ is the retarded propagator in its Wigner-transformed form \cite{Buss:2011mx}. Using the definition of the SF $\mathcal{P}$ as imaginary part of the propagator one can separate the spin-averaged spectral information from the phase-space content \cite{Kad-Baym:1962,Buss:2011mx}
\begin{equation}
F(x,p) = 2 \pi g f(x,p)\, \mathcal{P}(x,p)  ~,
\end{equation}
where $g$ is a spin--isospin degeneracy factor.
There is one KB equation for each particle, and all of them are coupled by the collision terms $C(x,p)$ and the mean-field potential in $H$. Setting the function $f \sim \sum_{i=1}^N \delta(\mathbf{x} - \mathbf{x_i(t)})\, \delta(\mathbf{p} - \mathbf{p_i}(t))$ then defines the trajectories of particles; this is the basis of the so-called test particle method used to solve the KB equations \cite{Bertsch:1988ik,Buss:2011mx}.

\paragraph*{BUU equations.} The structure of the KB equations can be simplified. Assuming on-shell particles, but still in a potential, one obtains \cite{Buss:2011mx}
\begin{equation}   \label{QPeq}
	F(x,p) = 2\pi g f(x,\mathbf{p})\, \delta(p_0 - E) ~,
\end{equation}
with $E = H = \sqrt{p^2 + {m^*}^2(\mathbf{x},\mathbf{p})}$ where for simplicity we assume that all potential effects are absorbed in a coordinate- and momentum-dependent effective mass (scalar potential).  Eq.\ \ref{QPeq} is the so-called quasi-particle approximation. With the potentials still present the KB equations become the so-called Boltzmann--Uehling--Uhlenbeck (BUU) equations. These still contain the all the effects of nuclear binding and the Pauli-principle.
%\begin{marginnote}[120pt]
%	\entry{BUU}{{\bf B}oltzmann--{\bf U}ehling--{\bf U}hlenbeck equation:  simplified version of the KB transport equations in which the potentials are kept within %a quasi-particle approximation}
%\end{marginnote}

The dissipative part of the transport is given by the so-called 'collision' term $C(x,p)$ in Eq.\ \ref{KBeq}. This term describes all the interactions of all the particles. In quasiparticle approximation, the collision term reads (dropping the $x$-dependence in $f$)\footnote{(The collision term is given here only for two-body collisions $p + p_2 \rightarrow p_1' + p_2'$, for the general case see \cite{Buss:2011mx}).}
\begin{widetext}
\begin{eqnarray}
C(x,p)
&=& \frac{g}{2} \int \frac{{\rm d}\mathbf{p}_2\,{\rm d}\mathbf{p}'_1\,{\rm d}\mathbf{p}'_2}{(2\pi)^9\,2E_p\, 2E_{p2}\, 2E_{p_1'} \,2E_{p_2'} }\, (2 \pi)^4 \delta^4(p + p_2 - p_1' - p_2')
\, \left\vert \mathcal{M}_{p p_2 \rightarrow p_1' p_2'}\right\vert^2 \nonumber \\
& & \mbox{}\times \left[ f(\mathbf{p}_1') f(\mathbf{p}_2' (1 - f(\mathbf{p}) (1 - f(\mathbf{p}_2)) - f(\mathbf{p}) f(\mathbf{p}_2) (1 - f(\mathbf{p}_1')) (1 - f(\mathbf{p}_2') )\right]
\end{eqnarray}
\end{widetext}
where $\mathcal{M}_{p p_2 \rightarrow p_1' p_2'}$ is the invariant matrix element for the transition.
The last line in this equation exhibits the familiar structure of a gain term and a loss term for the phase-space distribution of one particle, which is represented by $f(\mathbf{p})$. The factors $(1 - f(\mathbf{p}) (1 - f(\mathbf{p}_2))$ check whether the relevant phase-space region is unoccupied (Pauli-principle).

The transport equation, with its drift and collision terms, is Lorentz covariant. In practical applications the drift term respects relativity. Numerical algorithms have been developed to minimize any violations of relativity for the collision term \cite{Kodama:1983yk,Lang:1993,Kortemeyer:1995di}.

\paragraph*{Ground state properties.} At time $t = 0$ the phase-space distribution of nucleons is given by the Wigner transform of the ground-state one-particle density matrix, $\rho\left(\mathbf{x} - \mathbf{s}/2, \mathbf{x} + \mathbf{s}/2\right)$ :
\begin{equation}
f(\mathbf{x},0,\mathbf{p}) = \frac{1}{(2 \pi)^3} \int {\rm d}\mathbf{s}\,e^{-i \mathbf{p} \cdot \mathbf{s}} \rho\left(\mathbf{x} - \frac{\mathbf{s}}{2}, \mathbf{x} + \frac{\mathbf{s}}{2}\right)    ~.
\end{equation}
The density matrix can be obtained from, for instance, NMBT \cite{Benhar:2006wy}. In the simpler semi-classical theory, the ground-state distribution is given by the local Thomas--Fermi approximation
\begin{equation}
  f(\mathbf{x},0,\mathbf{p}) = \Theta\left(p_\mathrm{F}(\mathbf{x}) - |\mathbf{p}|\right)
\end{equation}
with the Fermi momentum $p_F(\mathbf{x}) \sim \rho(\mathbf{x})^{1/3}$. The hole SF is then given by
\begin{widetext}
\begin{equation}
\mathcal{P}_h(\mathbf{p},E) = g \int\limits_{\rm nucleus} \!\!\!{\rm d}^3x \, f(\mathbf{x},0,\mathbf{p}) \Theta(E)\, \delta\left(E - m^*(\mathbf{x},\mathbf{p}) + \sqrt{\mathbf{p}^2 + {m^*}^2(\mathbf{x},\mathbf{p})}\right)~.
\end{equation}
\end{widetext}
 The corresponding momentum distribution approximates that obtained in state-of-the-art NMBT calculations quite well (see figure 4 of Reference  \cite{Alvarez-Ruso:2014bla}); its energy distribution no longer contains the $\delta$-function spikes of a free Fermi gas because of the $\mathbf{x}$ dependence of the potential in $m^*$ and the integration over ${\rm d}^3x$.

\paragraph*{Inclusive cross sections}

The fully inclusive cross section is given by a sum over all possible subprocesses in the first time step where time $t = 0$ is defined as the moment of the first interaction of the incoming neutrino with a target nucleon. For the fully inclusive cross sections, further time development of the reaction is irrelevant. For example, for the QE contribution one has
\begin{equation}   \label{sigma}
{\rm d} \sigma^{\nu A}_{\rm QE} =  \int \frac{{\rm d}^3p}{(2\pi)^3} {\rm d}E\,\mathcal{P}_h(\mathbf{p},E) f_{\rm corr}\, {\rm d}\sigma^{\rm med}_{\rm QE} \, P_{\rm PB} (\mathbf{x},\mathbf{p})  ~.
\end{equation}
Here ${\rm d}\sigma^{\rm med}_{\rm QE}$ is a medium-dressed QE scattering cross section on a nucleon, $f_{\rm corr}$ is a flux correction factor $f_{\rm corr} = (k \cdot p)/(k^0p^0)$; $k$ and $p$ denote the four-momenta of the neutrino and nucleon momentum, respectively, and $P_{\rm PB}(\mathbf{x},\mathbf{p})$ describes the Pauli blocking. The inclusive cross section in Eq.\ \ref{sigma} agrees with that of the quantal theory.

\paragraph*{Fully exclusive cross sections.}
In order to obtain the fully exclusive final event in transport theory, the final-state particles of the very first initial interaction define the starting conditions for the next time step in the solution of the KB equation, and so on. With the production of particles, the number of equations effectively rises; with absorption it goes down. Therefore, the KB equations allow one to describe the inclusive cross sections consistently together with the exclusive ones. The calculation stops when the active particles are no longer interacting and then delivers the four-vectors of all of them.

\subsection{Event Simulation}
 Numerical solutions of the KB equations have been developed over the past 20 years. Among them GiBUU (Giessen--Boltzmann--Uehling--Uhlenbeck) is a consistent theory framework \cite{Buss:2011mx} and code \cite{gibuu}. It describes a wide class of nuclear reactions such as  ($A + A$), ($p,A$), ($\pi,A$), ($e,A$), ($\gamma,A$), and ($\nu,A)$ using the same physics input and code (see Reference \cite{Buss:2011mx}) for citations to these studies), and has been checked against many different nuclear reactions \cite{Leitner:2009ke}. It is consistent in the sense that it uses the same ground-state and collision dynamics for all processes, such as QE-scattering, pion production and DIS for neutrino-induced reactions.

 The ground state momentum distribution is given by the local Thomas--Fermi approximation. At the same time, the nucleons are bound in a coordinate- and momentum-dependent potential that has been fitted to equation of state and effective mass data \cite{Welke:1988zz,Gale:1989dm}. The single-particle cross sections discussed above are additional components,  and all the processes on the nucleus are assumed to be quasi-free. The two-particle two-hole (2p2h) component is -- similar to Reference \cite{Ivanov:2015aya} -- taken from an analysis of inclusive electron scattering data, namely the meson exchange current contribution in Reference \cite{Bosted:2012qc}. It can be related to the axial amplitude \cite{O'Connell:1972zz}. Furthermore, GiBUU has options for off-shell transport of hadrons with their in-medium SF and proper asymptotics. It is covariant and can thus be used also for high-energy collisions; for the collision term it uses the algorithms mentioned above \cite{Kodama:1983yk,Lang:1993,Kortemeyer:1995di}.

In addition to providing cross sections for many processes GiBUU produces full events and can thus be used as a neutrino event generator.

\paragraph*{Neutrino event generators.}
Standard neutrino event generators \cite{Andreopoulos:2009rq,Hayato:2009zz,Golan:2012rfa} are obtained from the KB theory after further simplifying assumptions, such as the quasi-particle approximation, the neglect of any potentials and any in-medium properties, all of them applicable at very high energies. Eq.\ (\ref{KBeq}) then becomes
\begin{equation}
\left(\partial_t + \frac{\mathbf{p}}{E} \cdot \mathbf{\nabla}_x\right) f(x,\mathbf{p}) = C(x,\mathbf{p}) ~.
\end{equation}
This equation forms the basis of all Monte Carlo event generators.
%\begin{textbox}[h]\section{NEUTRINO EVENT GENERATORS}
%Neutrino event generators have a twofold purpose. First, they are needed to take care of experimental details such as detector geometry for data evaluation. %Second, they are used to separate the signal of intest (e.g., QE-scattering) from other processes. Many groups also use them for comparison with models.
%\end{textbox}

In analyses of neutrino long- and short-baseline experiments neutrino-generators such as GENIE \cite{Andreopoulos:2009rq} and NEUT \cite{Hayato:2009zz} play a major role. They are needed to take care of experimen\-tal problems, such as target and flux geometry and experimental interfaces. They are also used to separate the signal from the background and thus have a direct influence on the final observables.  The quality of these observables is directly influenced by that of the generator. For a short review of generators see Gallagher \& Hayato in \cite{Agashe:2014kda}, and for a comparative discussion of various generators see Sect.\ 12.2 of Reference \cite{Leitner:2009zz}.

The neutrino generators use an ensemble of different theories to describe the various initial reaction processes \cite{Alam:2015nkk}. The FSI are then treated by Monte Carlo simulations based on phenomenological models. All nucleons are assumed to be free but Fermi-moving, and binding is taken into account only by correcting the final-state energies by a binding energy, usually one number. Some of the codes have a long history and often contain outdated physics  [e.g.\ the Rein--Sehgal form factors for resonances \cite{Rein:1980wg}]. Furthermore, the actual physics contents of these codes is not transparent, because there is usually no documentation of the physics and numerical algorithms used. Users of these generators often rely on tunes, namely parameter fits to observables, to make up for possible shortcomings in their physics content.
Tunes are often changed within the same experiment when different observables are being studied and they are often applied to separate pieces of the overall theory without concern for internal consistency. All of these factors limit the generators' predictive power for new targets or new energy regions.

\section{Lepton Interactions with Nuclei}
To understand the response of nuclei to incoming leptons one can use the large data base that arose from studies of the inclusive response of nuclei to incoming electrons \cite{Benhar:2006er}. These have shown that the components contributing to the total cross section on nucleons, specifically QE scattering, pion production, and DIS, also make up for most of the cross section on nuclei. Both the experimental and the theoretical developments in this field until the early 1990s are comprehensively discussed in the textbook by Boffi et al. \cite{Boffi:1994}. Inclusive cross sections obviously constitute a necessary test for any model description. From an experimental point of view they are 'clean' because they require generator use only for the truly experimental problems, thereby minimizing any model dependence.

\subsection{Electron Interactions with Nuclei}
For QE scattering both the nuclear ground-state structure and the reaction mechanism and potential felt by the outgoing nucleon are essential. The impulse approximation (quasi-free interaction with one nucleon at a time) is expected to be reliable for momentum transfers larger than approximately 300 MeV \cite{Ankowski:2011dc}. One then expects that inclusive total cross sections should scale linearly with target mass number $A$. Sealock et al.\ \cite{Sealock:1989nx} have explored targets from helium to tungsten. The scaling $\sim A$ indeed held quite well for the $\Delta$ resonance region and was slightly disturbed ($\approx 10\%$) for the QE peak. The missing strength in the dip region between QE peak and $\Delta$ was explained by processes in which the incoming photon interacts not with just one nucleon, as for true QE interactions, but with two nucleons simultaneously \cite{Alberico:1983zg,Dekker:1994yc,VanderSluys:1995rp,DePace:2003xu}.

The Valencia and Torino groups developed a comprehensive description of the nuclear electromagnetic response by starting from a local Fermi gas in a diagrammatic approach \cite{Gil:1997bm,Gil:1997jg,Alberico:1997jg}. The calculation included random phase approximation (RPA) excitations and thus took care  of the most important collective modes in a Fermi gas.

An alternative method is the scaling approach \cite{Day:1990mf}, in which electron data are used to extract a universal scaling function that can then be used to calculate cross sections for different kinematical regimes and even different target mass numbers \cite{Antonov:2011bi}. The method relies on the assumption of an independent particle model. Thus, effects of 2p-2h excitation or of inelastic excitations have to be added manually, either by calculating such contributions \cite{Amaro:2010iu,Megias:2014qva} or by taking them from parameterizations of earlier results \cite{Ivanov:2015aya}.

%\begin{marginnote}[180pt]
%	\entry{NMBT}{Nuclear Many Body Theory}
%\end{marginnote}
NMBT, pioneered in the 1990s, made it possible to determine the nuclear ground state SFs in Eq.\ \ref{sigma}, starting from a realistic nucleon-nucleon interaction \cite{Benhar:1994hw,Benhar:2006wy}. The SF method has been applied to QE scattering; for that it has to be supplemented by a model for the interactions of the scattered nucleon in the final state \cite{Ankowski:2014yfa}. The SFs also carry information about the so-called short-range correlations (SRC) between nucleon pairs that are responsible for the high-momentum tails; the latter are missing in the Fermi gas-based models. Differences between the predictions from the local Fermi gas and the NMBT SFs are to be expected only in experiments particularly sensitive to these high-momentum tails.

Recently, ab initio Quantum Monte Carlo calculations became available that give an excellent description of nuclear ground and excited states, at least for light nuclei up to $^{12}$C \cite{Carlson:2014vla}. These calculations have already been employed in a calculation of the inclusive electroweak response of these nuclei and have yielded interesting insight into the importance of 2p-2h processes \cite{Lovato:2015qka}.

All of the methods discussed above are suitable for calculation of inclusive cross sections only. None of them can handle high-multiplicity final states. These final states can be handled only by the transport-theoretical method for photon- and lepton-induced reactions \cite{Buss:2011mx}. For inclusive cross sections the results of the latter are compatible with those from the Valencia and Torino groups \cite{Effenberger:1999jc,Lehr:1999zr,Buss:2007ar,Leitner:2008ue,Gallmeister:2016dnq}.

\subsection{Neutrino Interactions with Nuclei}
Extracting the various components, QE scattering, pion production, and DIS, from neutrino-induced data on nuclei is significantly more complicated than for electrons. The energy distribution of a neutrino beam is broad, and the incoming neutrino energy, and, therefore, also the momentum- and energy-transfer, must be reconstructed from observations of the final state. Furthermore, the identification of a particular reaction mechanism, such as QE scattering, is necessarily model dependent. The final state of a true QE reaction and of a pion production event, with subsequent pion absorption inside the nuclear target, are experimentally indistinguishable. Thus, the 'data' for QE scattering can never be better than the methods used to describe pion production on nuclei.
%\begin{textbox}[h]
%	\section{Quasielastic Scattering}
%    QE Scattering on nuclear targets is experimentally indistinguishable from pion production with subsequent pion absorption. The data for QE scattering can %never be better than the theory used to describe pion production and absorption on nuclei.
%\end{textbox}

For comparison with experimental neutrino data, the fixed-energy cross section described above is folded with the appropriate flux (neutrino energy distribution) $\phi(E_\nu)$,
\begin{equation}
\langle d\sigma \rangle = \int {\rm d}E_\nu \, \phi(E_\nu) \,{\rm d}\sigma^{\rm med}(E_\nu) ~,
\end{equation}
where $\phi(E_\nu)$ is the incoming energy distribution (the flux), normalized to one, and ${\rm d}\sigma^{\rm med}(E_\nu)$ is the cross section on the nuclear target, with appropriate medium corrections.

\subsubsection{Quasi-elastic scattering} 
\begin{figure}
\includegraphics[width=3in]{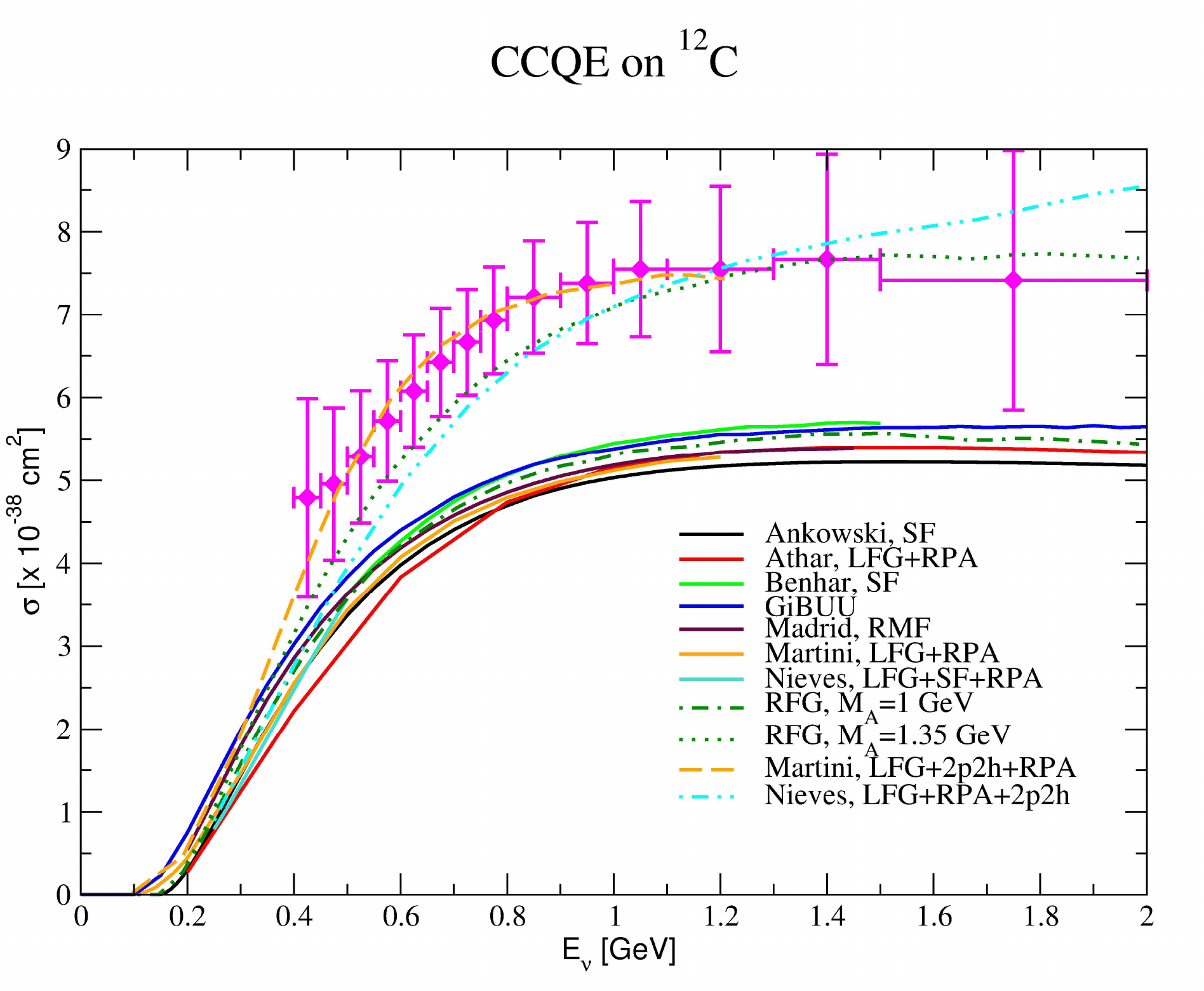}
\caption{CCQE cross section for a CH$_2$ target obtained in the MiniBooNE experiment \cite{miniboone}. The many lower curves give various theoretical predictions for the true QE events, obtained with an axial mass of 1.03 GeV; the dotted green curve gives the result for true QE events obtained with an increased axial mass of 1.3 GeV. The two dashed curves give predictions of models that take many-body interactions into account \cite{Martini:2009uj,Nieves:2011pp}. (from \cite{Alvarez-Ruso:2014bla}).}
\label{fig:MBQE}
\end{figure}

The same methods summarized for electron scattering in the preceding sections have been applied also to the description of QE events in neutrino-nucleus reactions \cite{Leitner:2006ww,Leitner:2006sp,Carlson:2014vla,Delorme:1985ps,Marteau:1999jp,Kolbe:2003ys,Nieves:2004wx,Benhar:2006nr,Ivanov:2013saa,Ivanov:2015aya}.
QE scattering was assumed to be well understood in terms of interactions of the incoming gauge boson with single nucleons, and data on nuclear targets were indeed reproduced in this framework once so-called 'stuck-pions' events (i.e., events in which a pion was first produced but later reabsorbed) were removed by generators. For example, the NOMAD experiment \cite{Lyubushkin:2008pe}, working with a target mainly consisting of carbon, extracted an $M_{\rm A}$ value of 1.05 GeV, in agreement with the world average value of 1.03 GeV (see Reference \cite{Lyubushkin:2008pe} for a table showing results from other experiments).  Therefore, it came as a surprise that experiments using large-volume Cherenkov counters (K2K, MiniBooNE) \cite{Gran:2006jn,:2007ru} found a cross section well above the model predictions for QE  (\textbf{Figure \ref{fig:MBQE}}) \footnote{Neither the cross section nor the neutrino energy on the axes in \textbf{Figure \ref{fig:MBQE}} is directly observable. Both have been reconstructed with the help of a particular neutrino generator.}. The measured higher cross sections required an $M_{\rm A}$ value of about 1.3 GeV, which is significantly larger than the world-average value of 1.03 GeV.

Around 30 years ago Delorme \& Ericson \cite{Delorme:1985ps} realized that in certain detector types 2p-2h excitations could be experimentally indistinguishable from true QE events and would thus contribute to the QE cross section (see also Reference \cite{Marteau:1999jp}).%
%\begin{marginnote}[120pt]
%	\entry{2p-2h}{2particle-2hole processes in which the incoming neutrino interacts with 2 nucleons simultaneously, often connected with the excitation of one %nucleon to its $\Delta$ resonance}
%\end{marginnote}
This fact was 'rediscovered' by Martini et al.\ \cite{Martini:2009uj,Martini:2010ex,Martini:2011wp} who pointed out that a consideration of 2p-2h initial interactions, taken together with RPA excitations of the nucleus, could explain not only the observed energy dependence depicted in \textbf{Figure \ref{fig:MBQE}} but also the measured double-differential cross sections for these QE events without increasing $M_{\rm A}$. In a related model Nieves et al.\ \cite{Nieves:2011pp,Nieves:2011yp} pursued the suggestion of explaining the MiniBooNE surplus cross section by 2p-2h excitations.

The experiment MINER$\nu$A has also attempted to extract experimental information on 2p-2h contributions in another, higher-energy range of a few GeV \cite{Fiorentini:2013ezn,Walton:2014esl} by analyzing the ${\rm d}\sigma/{\rm d}Q^2$ distributions. The results of these investigations are inconclusive \cite{Megias:2014kia} partly because (a) a large pion background has to be subtracted (see the discussion in Sect.\ \ref{s:nuclpionprod}) (b) $Q^2$ cannot be directly measured but rather has to be reconstructed with large errors just in the relevant region of largest cross sections \cite{Mosel:2014lja}, and (c), the experiment observes outgoing muons only under forward angles where the relative effect of 2p-2h processes on QE-like events is smallest.

A more detailed discussion of neutrino-induced QE scattering can be found in recent reviews \cite{Gallagher:2011zza,Alvarez-Ruso:2014bla}.

\subsubsection{Final-state interactions}
The analysis of long-baseline neutrino experiments, to be discussed in Sect.\ \ref{s:LBL}, requires a description of the full final state. That necessitates not only a description of the initial, primary interaction of the neutrino with a bound and Fermi-moving nucleon leading to all sorts of final states, but also a reliable description of the FSI that the initially produced hadrons experience on their way out of the nuclear target, both with other target nucleons and among themselves. A consistent theory should be able to describe both the inclusive and the exclusive events.

The only directly measurable observables from a neutrino-nucleus reaction are the momenta of outgoing leptons, nucleons and mesons. In the following subsections I discuss emitted nucleons and produced pions. Outgoing leptons are treated implicitly in the above discussion of QE cross sections.

\subsubsection{Knockout nucleons}

\begin{figure}
\includegraphics[width=3.5in]{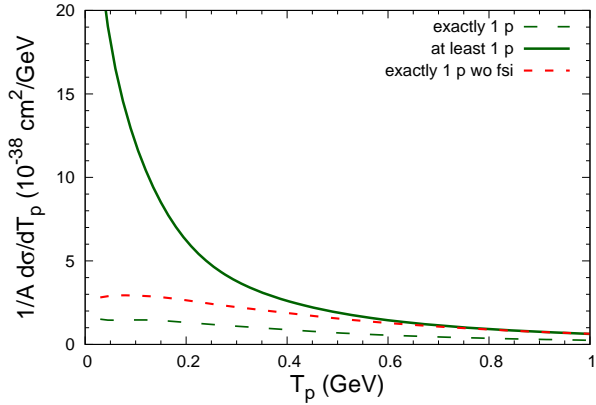}
\caption{Kinetic energy spectrum of knock-out protons in the MINER$\nu$A experiment with an average neutrino energy of 3.4 GeV on a $CH$ target. The short-dashed red curve represents the spectrum of events with exactly one outgoing proton without FSI, the dashed green curve gives the same spectrum with FSI and the topmost solid green curve describes the semi-inclusive spectrum of 1 proton after FSI (from \cite{Mosel:2014lja})}
\label{fig:Nknockout}
\end{figure}
One of the observables most sensitive to FSI is the spectrum of emitted nucleons, depicted in \textbf{Figure \ref{fig:Nknockout}}. The figure shows the cross section for (exactly) one-proton events before FSI. Once the FSI become active, this spectrum is suppressed for all the kinetic energies. The reason for this suppression is the so-called 'avalanche effect', in which the initially produced proton collides with other nucleons and ejects more and more protons. Energy conservation then requires that these secondary protons have lower energies, as shown in the figure by the curve that steeply rises towards smaller proton kinetic energies for semi-inclusive 1p events. The steepness of this pileup at small kinetic energies demonstrates that the total number of protons detected depends sensitively on experimental detection thresholds.  A description within the optical model or the multiple scattering theory describes only the flux loss at a given kinetic energy; it does not provide any information where the absorbed flux goes \cite{Martinez:2005xe,Meucci:2004ip}.

At neutrino energies above approximately 1 GeV the knockout nucleons come, with approximately equal probability, from true one-body QE scattering and an initial $\Delta$ production followed by the pionless $\Delta$ decay $\Delta N \rightarrow N N$ \cite{Leitner:2006ww}. Naively, the investigation of two-nucleon knockout could signal the presence of these many-body interactions. However, a closer analysis has shown that the shape of the kinetic energy distributions is not changed by the presence of 2p-2h interactions \cite{Lalakulich:2012ac}. Even in a subset of events with only two outgoing nucleons there are many events due to initial one-body interactions (because of the avalanche effect).

\subsubsection{Pion Production}   \label{s:nuclpionprod}

\begin{figure}[!hbt]
	\begin{minipage}[c]{0.48\textwidth}
		\includegraphics[width=\textwidth]{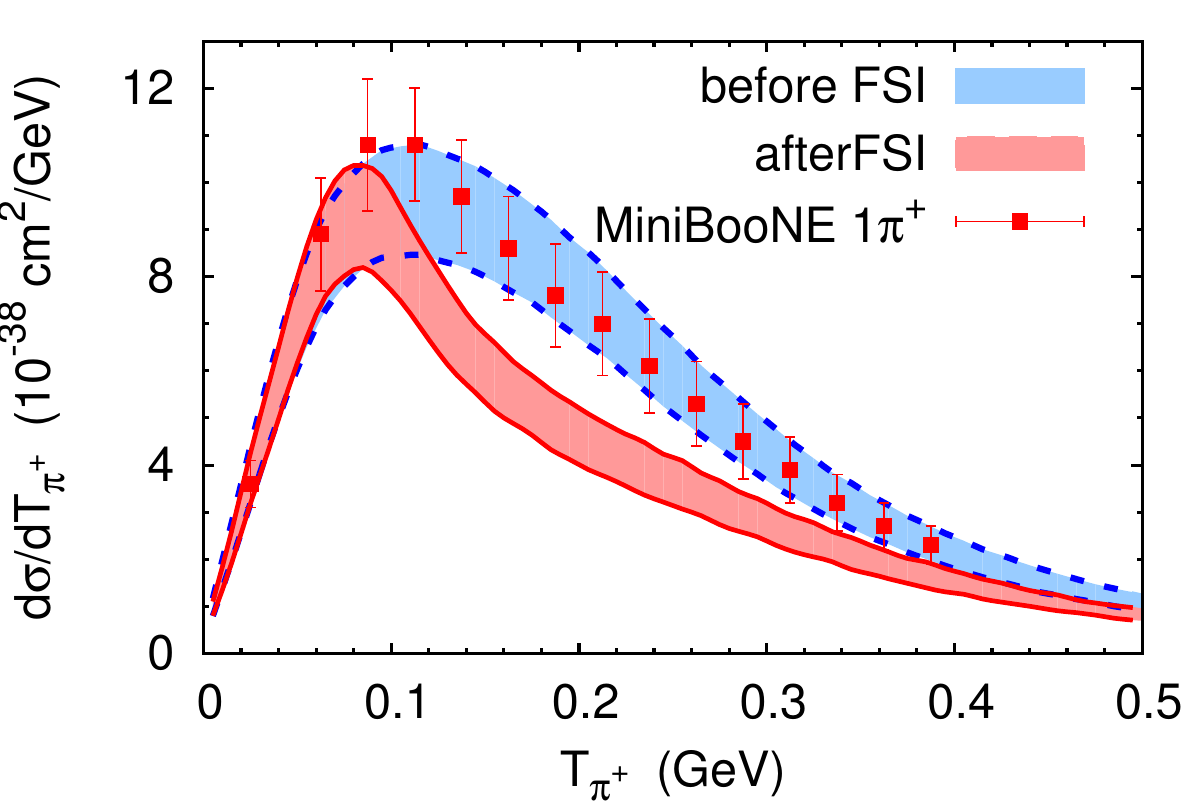}
	\end{minipage}
	\hfill
	\begin{minipage}[c]{0.48\textwidth}
		\includegraphics[width=\textwidth]{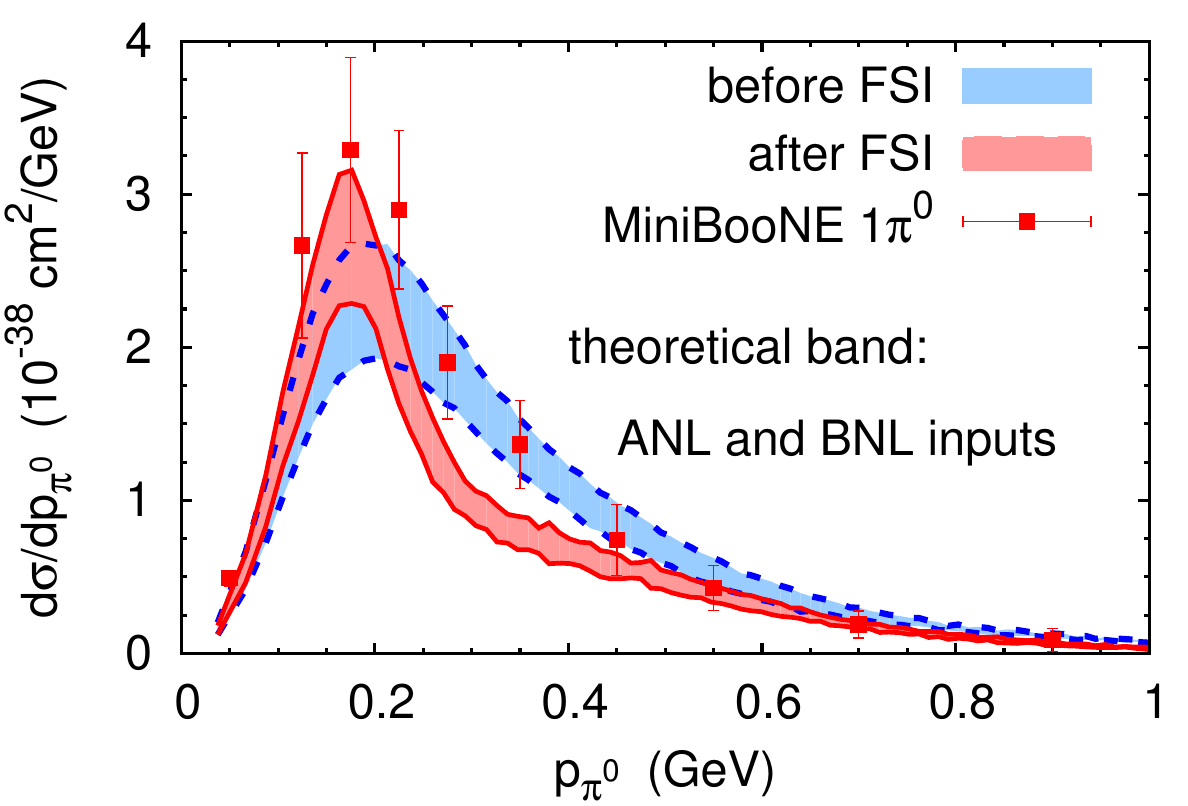}
	\end{minipage}
	\caption{Kinetic energy distribution of the outgoing $\pi^+$ and momentum distribution of the outgoing $\pi^0$ for one-pion production at MiniBooNE. Data are from~\cite{AguilarArevalo:2010bm,AguilarArevalo:2010xt}. The upper (blue) band shows the results obtained from a GiBUU calculation without any FSI on the pion; the upper border of the band corresponds to the BNL input, the lower to the ANL input. The lower (red) band corresponds to the GiBUU results after all FSI have been turned on. (from \cite{Lalakulich:2013tba})}
	\label{fig:MB-pion-dTkin}
\end{figure}

The first extensive data set on pion production was obtained by the MiniBooNE experiment \cite{AguilarArevalo:2010bm,AguilarArevalo:2010xt}; the data are shown in {\bf Fig.\ \ref{fig:MB-pion-dTkin}}.
	
Theoretically, Leitner et al.\ \cite{,Leitner:2006ww,Leitner:2006sp} studied pion production on nuclei by using the impulse approximation on a nuclear ground state, with a local Fermi gas momentum distribution in a mean-field potential. The cross section in Eq.\ (\ref{sigma-pi}) was evaluated in the rest frame of each bound, Fermi-moving nucleon and GiBUU was used to describe the all important FSI.

In-medium effects for pion production are contained both in the SF $\mathcal{A}$ of the initially excited nucleon resonances (Eq.\ \ref{resprod}) and in the branching ratio for the resonance decay into a pion and a nucleon, where the final nucleon state may be Pauli blocked. Both the initial-state and the final-state nucleons are bound in a momentum- and coordinate-dependent potential that -- through energy- and momentum-dependence -- affects the decay width \cite{Oset:1987re}. The theory has been tested with the help of photon-induced \cite{Krusche:2004uw} and electron-induced \cite{Kaskulov:2008ej} pion production data.

Most of the pions at the energies of the MiniBooNE experiment, with its flux peak at a neutrino energy of approximately 600 MeV originate from the $\Delta$ resonance. The high-energy tails of the incoming flux add some minor contributions from higher resonances and DIS \cite{Lalakulich:2012cj}. A small amount of pions also comes from the initial QE vertex, when the outgoing proton is rescattered. Here the main contribution is from the $p N \to N' \Delta \to N' N^{''} \pi$ reaction. Other possibilities to create pions during the FSI would be $\omega N \to \pi N$, $\phi N \to \pi N$, $\pi N \to \pi \pi N$.

FSI noticeably decrease $\Delta$-originated pion production due to the absorption $N\Delta\to NN$; a similar process is also possible for other resonances. Once a pion is produced, independently of its origin, it may also undergo a charge exchange, $\pi^+ n \to \pi^0 p$, which depletes the $\pi^+$ channel as the dominant one, but increases the $\pi^0$ channel. The latter effect at a momentum of approximately 180 MeV is depicted in {\bf Fig.\ \ref{fig:MB-pion-dTkin}} (right), which shows that the cross section after FSI is larger than that before. Other possible ways for pions to disappear, at higher energies, include $\pi N \to \omega N$, $\phi N$, $\Sigma K$, $\Lambda K$.

{\bf Fig.\ \ref{fig:MB-pion-dTkin}} shows a significant disagreement between theory and data, in both the magnitude and shape of the kinetic energy distribution. The figure presents results obtained with GiBUU \cite{Lalakulich:2012cj}; an independent calculation by Hernandez et al.\ \cite{Hernandez:2013jka} yields essentially the same result. The shape is a direct consequence of the well-understood $\pi N \Delta$ dynamics in nuclei. Pions with a kinetic energy of approximately 200 MeV are strongly reabsorbed through the sequence of reactions $\pi N \rightarrow \Delta$ and $\Delta N \rightarrow N N$. This shape has been experimentally observed in the kinetic energy distributions of $\pi^0$ produced in reactions with photons in the energy regime up to approximately 1 GeV on nuclei \cite{Krusche:2004uw}. Since the FSI are the same in both reactions, the neutrino spectra should exhibit the same behavior.

Other data on pion production on nuclear targets are still sparse. The MINER$\nu$A experiment, operating with a flux that peaks at a considerably higher energy of around 3 GeV, obtains cross sections \cite{Eberly:2014mra,Aliaga:2015wva} that are close to the ANL input \cite{Mosel:2015tja}. The shape of the distributions is closer to that expected from theory. There is a conflict with the MiniBooNE data, however, as far as the absolute cross section is concerned \cite{Sobczyk:2014xza}
%\begin{textbox}[h]
%	\section{Pion Production}
%	Neutrino-induced pion production, through resonances or DIS, is the major part (~ 2/3) of the total cross section at higher energy long-baseline experiments, %but is not well under control.
%\end{textbox}

In summary, neutrino-induced pion production is still not well understood. This is disturbing, given that pion production makes up most of the background for QE scattering events. In the MiniBooNE and T2K energy range pion production accounts for approximately one third and in the MINOS, MINER$\nu$A and DUNE range it accounts for approximately two thirds of the total cross section \cite{Mosel:2012kt,Mosel:2015oda}. The soon to be released data on pion production from T2K may shed some light on this puzzle \cite{Lalakulich:2013iaa}.

MINER$\nu$A also has plans to explore the weak strangeness production process on nucleons \cite{Solomey:2005rs}. The very strong pion FSI make this plan rather difficult. GiBUU simulations have shown that most of the kaons produced originate in secondary processes such as $\pi N \to \Lambda K$ \cite{Lalakulich:2012gm,Mosel:2014lja}. Although kaons will undoubtedly be produced, they are mostly not the ones from an initial neutrino-induced reaction.

\section{EFFECTS OF NUCLEAR INTERACTIONS IN LONG-BASELINE PHYSICS} \label{s:LBL}
\subsection{Energy Reconstruction}
In long-baseline experiments searching for neutrino oscillations, such as T2K, MINOS, NOvA and the future DUNE (formerly called LBNE), the event rate at a given neutrino energy $E_\nu$ for a far detector is compared with that at a near detector. At both detectors the neutrino energy must be reconstructed event by event from the final state of the reaction.

Two methods for this energy reconstruction are being considered:

\begin{enumerate}
\item
In the so-called calorimetric method the energy of the final-state particles is observed. This is the method that will be used in the liquid argon detectors. It requires an accurate determination of the final-state energy. If the detector were perfect, it would directly provide the incoming beam energy, through energy conservation. However, because of acceptance limitations, actual detectors observe only a part of the energy of the final-state particles and must extrapolate from that to the full final-state energy. Initial studies have shown that the effects of experimental detection thresholds on the reconstruction can be quite large \cite{Leitner:2010kp,Ankowski:2015jya}.

\item
For QE charged-current scattering of a neutrino on a free nucleon at rest the incoming neutrino energy can be determined completely from the outgoing lepton kinematics (energy and angle). This is the method used largely by lower energy experiments, such as MiniBooNE and T2K. It requires a correct identification of the reaction mechanism as being QE scattering. Fermi-motion of bound nucleons alone leads to a smearing of the reconstructed energy around the true value, with an uncertainty width of approximately 60 MeV for a neutrino energy of approximately 1 GeV \cite{Leitner:2010kp}. This is a natural lower limit to the error with which the neutrino energy can be reconstructed in this method. Errors introduced through an incorrect identification of a QE scattering event are even larger. An example is the misidentification of 2p-2h events as QE scattering that led to the extraction of unphysical values for the axial mass from experiments with nuclear targets \cite{:2007ru,Adamson:2014pgc,Abe:2015oar}.
\end{enumerate}

The quality of both of these methods could be assessed by performing analyses of model-generated events. Leitner \& Mosel \cite{Leitner:2010kp} generated events with GiBUU and found that the presence of stuck-pion events always leads to a low-energy tail on the distribution of reconstructed energies. The same is true for the presence of 2p-2h events \cite{Lalakulich:2012hs,Martini:2012fa,Martini:2012uc,Nieves:2012yz,Benhar:2013oba}. Additionally, other reaction mechanisms, such as DIS, lead to such a reduction of the reconstructed energy \cite{Lalakulich:2012hs}.

\subsection{Energy Reconstruction and Oscillations}
The difficulties reconstructing the incoming energy also affect the oscillation signal.

\subsubsection{T2K}
In the T2K energy regime, where the flux peaks around 600 MeV, pion production accounts for approximately one third of the total cross section \cite{Mosel:2012kt} and the stuck-pion events are nearly as frequent as the 2p-2h events \cite{Lalakulich:2012hs}.
%\begin{marginnote}[120pt]
%\entry{T2K}{Long-baseline experiment from Tokai to Kamiokande in Japan}
%\end{marginnote}
It is then interesting to determine how the errors due to the misidentification of events as being QE affect the oscillation signal. {\textbf{Fig.\ \ref{fig:T2Kosc}}(left) the results of such a study.
\begin{figure}[!hbt]
\begin{minipage}[c]{0.48\textwidth}
\includegraphics[width=\textwidth]{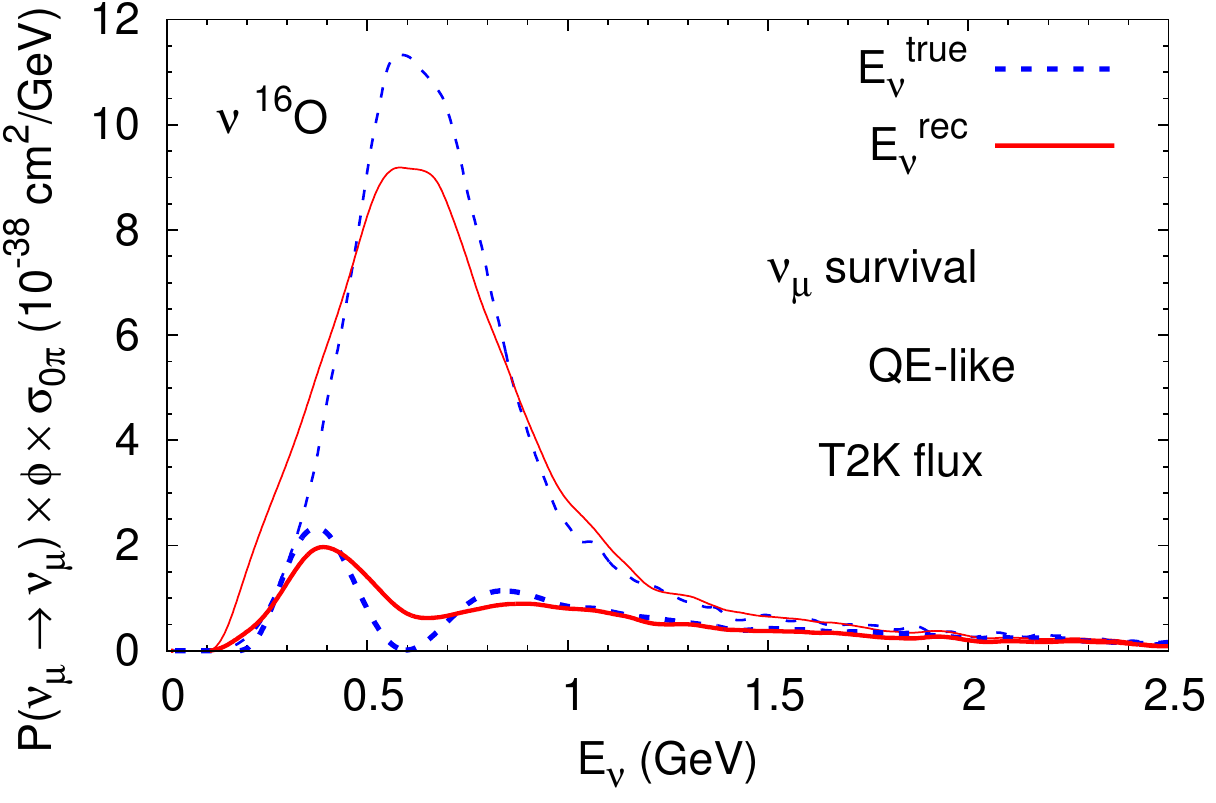}
\end{minipage}
\hfill
\begin{minipage}[c]{0.48\textwidth}
\includegraphics[width=\textwidth]{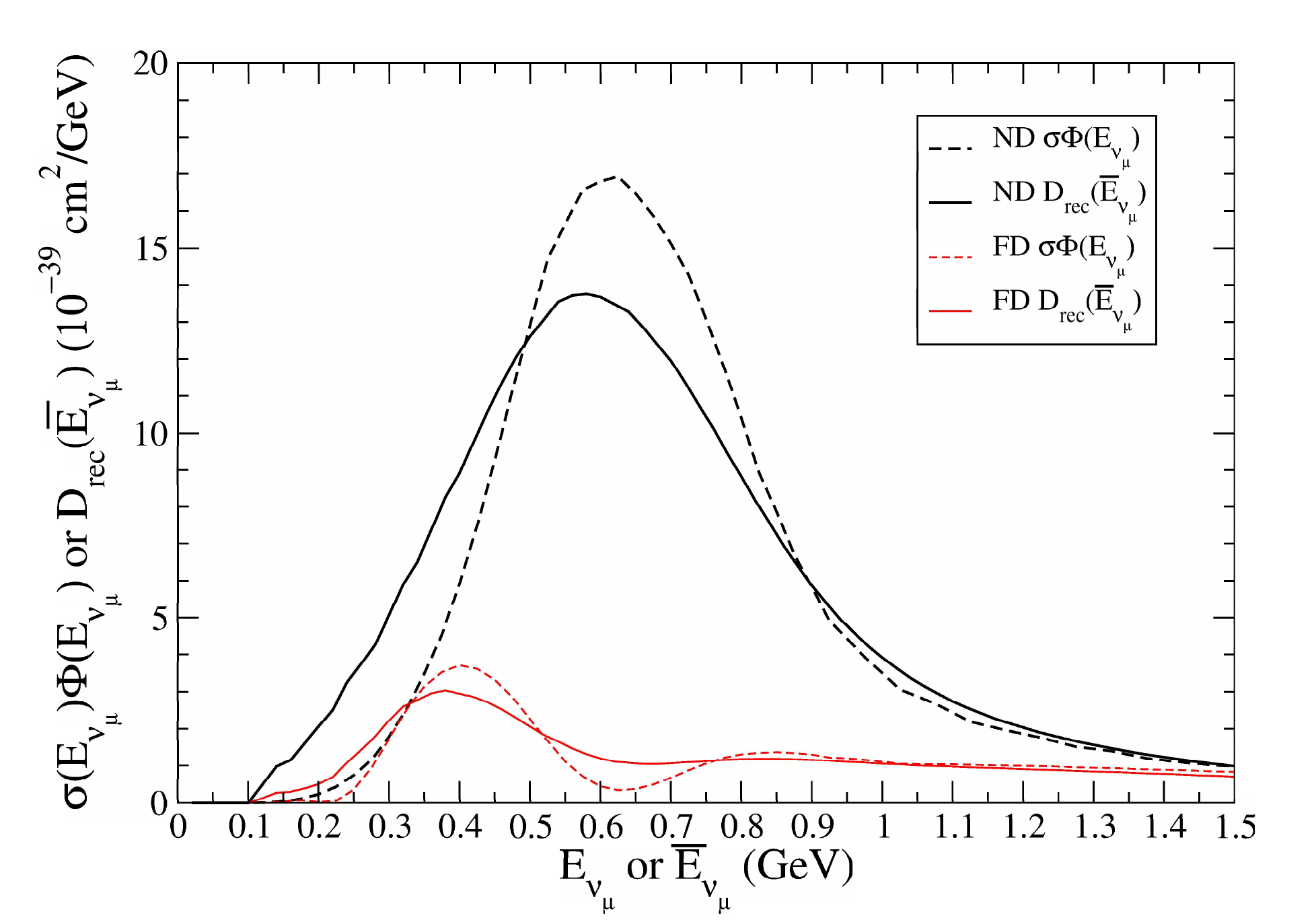}
\end{minipage}
\caption{Muon neutrino survival rates at the near ({\it upper curves}) and far ({\it lower curves}) detectors at the T2K experiment. The dashed lines give represent the distributions as function of the true neutrino energy; the solid lines represent those as function of reconstructed energies. The left figure shows results obtained with GiBUU and is taken from \cite{Lalakulich:2012hs}, the right shows results of a RPA-based many-body calcultion and is taken from \cite{Martini:2012uc}.}
\label{fig:T2Kosc}
\end{figure}
The left part of that figure shows that the reconstructed signal at the near detector is again shifted to lower energies and that the oscillation minimum at around 600 MeV is smeared out and thus harder to locate.

Exactly the same behavior is also found in very different calculations by Martini et al. \cite{Martini:2012uc}  ({\textbf{Fig.\ \ref{fig:T2Kosc}(right)}). Both the shift to lower energies in the near signal and the smearing of the oscillation signal agree very well with these features in the right part of {\textbf{Fig.\ \ref{fig:T2Kosc}}).

The oscillation signal here was obtained by using some reasonable values for the mixing angles. The authors of \cite{Coloma:2013tba} went one step further by directly looking on the effects of the energy reconstruction on the extracted oscillation parameters. For this study, events were generated with GiBUU (assumed to represent the measured data, i.e.\ 'nature') and then analyzed with the widely used neutrino generator GENIE \cite{Andreopoulos:2009rq}. The authors found that both the mixing angle and the squared mass difference changed.

\subsubsection{DUNE}
At the higher energies of the NuMI beam at Fermilab and at the planned LBNF/DUNE experiment, which peaks at approximately 3.5 GeV, pion production is the dominant component. Studies of the remainder, QE, then require both a quantitative understanding of pion production, both through resonances and through DIS, and a reliable implementation of that understanding in generators.

%\begin{marginnote}[-140pt]
%\entry{DUNE}{Deep Underground Neutrino Experiment}
%\end{marginnote}
{\bf Fig.\ \ref{fig:mue-app}} illustrates the impact of the QE-based reconstruction on the oscillation signal, plotted for two different event samples as a function of true neutrino energy and of energy reconstructed from the outgoing electron kinematics assuming a true QE process. Not only do errors in the energy reconstruction due to event misidentification cause a shift of the energy axis, they distort the whole event distribution \cite{Lalakulich:2012hs,Martini:2012fa}.
\begin{figure}[!ht]
\begin{center}
\includegraphics[width=0.9\columnwidth]{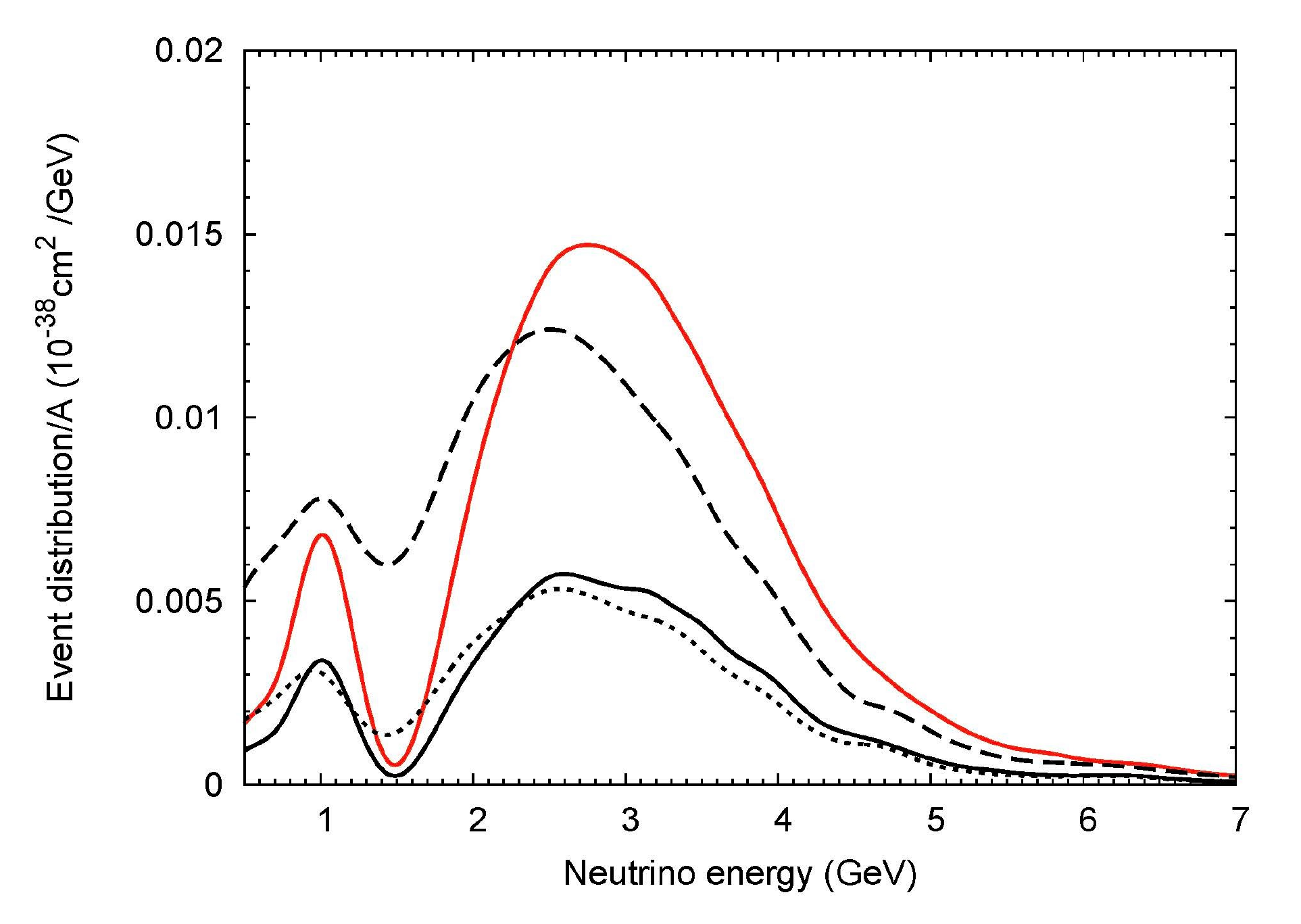}
\caption{Distribution of $\nu_e$ appearance events (normalized flux times cross section) per nucleon for DUNE vs.\ true ({\it solid curve}) and reconstructed ({\it dashed curve}) energy. The upper two curves show the results obtained from an event sample with zero pions, the two lower curves are obtained from a sample with zero pions, one proton and $X$ neutrons (\cite{Mosel:2013fxa}), showing the expected event distribution for electron appearance at DUNE.} \label{fig:mue-app}
\end{center}
\end{figure}
Also shown in the figure is the oscillation signal obtained from an event sample with zero pions that was produced by GiBUU; the energy was then reconstructed using the QE-based method \cite{Mosel:2013fxa}. The event rate vs.\ reconstructed energy (upper dashed curve) is distorted in comparison to the event rate  vs.\ true energy and is shifted by more than 500 MeV in its maximum. This is clearly above the accuracy required to distinguish between the various parameter scenarios in Fig.\ \ref{fig:LBNE-oscill}.

The situation is significantly improved when the event sample is further restricted to contain one and only one proton (plus any number of neutrons). Now the difference in energy between the true and the reconstructed curves is at most 100 MeV. The fact that the energy reconstruction is based on the dynamics of a true (one-body) process implies that requiring one proton in addition to zero pions gives a significantly cleaner identification of true QE because it singles out events that occur in the nuclear surface with less rescattering.  This event selection has recently also been attempted in a study of QE scattering by the MINER$\nu$A experiment \cite{Walton:2014esl}.

The same improvement also occurs for the difference between the true and reconstructed oscillation signals in dependence on $\delta_{CP}$ \cite{Mosel:2013fxa}. Experiments searching for this phase would be well advised to look at events with one lepton, zero pions, one proton and $X$ (unobserved) neutrons. With this subsample, the QE-based energy reconstruction should also be sufficiently reliable at the higher energies of the DUNE experiment. It could provide a useful alternative to the calorimetric method, which is also plagued by uncertainties \cite{Ankowski:2015jya}.

How uncertainties in the generators used actually affect the oscillation mixing angles and other neutrino properties has been discussed in recent papers \cite{Coloma:2013rqa,Coloma:2013tba,Ankowski:2015jya}.

\section{SUMMARY}
Neutrino interactions with \emph{nucleons} are not well known. QE scattering still suffers from large experimental uncertainties, which translate directly into uncertainties in the shape of the axial form factor. For the first inelastic process, pion production through the $\Delta$ resonance, this is even more so; there a larger number of form factors is essentially unconstrained by presently available data. Contributions from higher nucleon resonances to pion production are constrained only in their strength by PCAC. Only at very high neutrino energies, above approximately 30 GeV, does DIS become the relevant reaction channel and this is, being a pQCD processs, on safer grounds. Unfortunately, from a theoretical point of view, all planned long-baseline neutrino experiments work with neutrino energy distributions that peak at a few GeV --- that is, in the theoretically extremely challenging region between resonance physics and DIS. Gaining a more precise understanding of the neutrino-nucleon cross sections in the region below DIS requires new data with the elementary targets hydrogen and deuterium.

Neutrino interactions with \emph{nuclei} are interesting from the point of view of many-body physics. Their practical importance stems from the fact that the targets in modern ongoing (T2K, NOvA, MINER$\nu$A, MicroBooNE) \cite{Rubbia:2016} or planned (DUNE) experiments \cite{Acciarri:2015uup} are all nuclei such as carbon, oxygen and $^{40}$Ar. From these experiments neutrino oscillation parameters can be extracted only if the incoming neutrino energy is known. This energy has to be reconstructed from final state particles, requiring knowledge of neutrino-nucleon interaction rates in medium and of the FSI of the outgoing hadrons.  Therefore, it is encouraging that a broad experimental program including the MiniBooNE, MicroBooNE, MINER$\nu$A experiments and the T2K near detector, is dedicated to  measuring neutrino interaction cross sections

These experiments have to rely on event generators for taking care of various experimental effects, such as detector and flux geometry. Generators are often used to describe the data as well, thereby replacing a consistent theoretical analysis even though they often lag behind in their implementation of present-day nuclear physics.

A future challenge will be to bring the generators into a closer relationship with a theory that is able to describe the complete time-development of the neutrino-nucleus reaction with all processes included; inclusive cross sections or QE interactions alone are not enough. There has been tremendous progress in the theory of lepton-nucleus interactions. Ground-state properties can now be calculated from first principles with much higher accuracy than ever before and reaction mechanisms are becoming better and better understood. Equally important, the description of the dynamical evolution of the nuclear system has benefitted from the development of by now well-established quantum-kinetic transport-theoretical methods in other fields of physics.  Employing these state-of-the-art theoretical methods of transport theory and nuclear physics is essential for extending the event generators into new regions of energy and target mass. The precision era of neutrino physics also requires new, precision-era generators.

\subsection{Summary Points} A short overall summary then reads as follows:
\begin{enumerate}
	
	\item Cross sections for neutrino interactions with nucleons suffer from large experimental uncertainties, both for QE scattering and $\Delta$ resonance excitations. These elementary cross sections enter into the description of neutrino interactions with nuclei.
	
	\item Cross sections for neutrino interactions with nuclei offer access to the electroweak response of nuclei. Pion production is the dominant reaction component at higher energies. All studies of QE scattering are limited by the accuracy with which pions can be described.
	
	\item Many-body reaction mechanisms connected with 2p-2h excitations in the target nucleus play a role also in neutrino induced reactions.
	
	\item Oscillation parameters can be extracted from long-baseline experiments only with the help of neutrino event generators. Generators, therefore, play an all-important role, in contrast to most other experiments in nuclear and hadron physics.
	
	\item The QE-based energy reconstruction offers a viable alternative to the calorimetric method also at higher energies if the proper event samples (1 $\mu$, 0 $\pi$, 1 $p$, $X$ $n$) are chosen.
	
\end{enumerate}

\subsection{Future work}
Future work in this field should follow the following points:
\begin{enumerate}
\item New, more precise experimental determinations of cross sections on elementary targets ($p, D$) are needed to minimize uncertainties in the description of neutrino-nucleus interactions.

\item NMBTs for the electroweak response of nuclei have to be extended to noninclusive event descriptions and inelastic processes.

\item The use of neutrino event generators should be kept to a minimum. Published data should contain as little generator dependence as possible.

\item  More accurate theoretical analyses of measured event rates, for extracting interaction cross sections or neutrino oscillation parameters, have to be developed. They should be  based on state-of-the-art methods of nuclear physics, not only for static nuclear structure but also for nuclear reactions.

\end{enumerate}

%Disclosure
%\section*{DISCLOSURE STATEMENT}
%The author is not aware of any affiliations, memberships, funding, or financial holdings that
%might be perceived as affecting the objectivity of this review.

% Acknowledgements
\section*{ACKNOWLEDGMENTS}
I gratefully acknowledge the support of the whole GiBUU group and many stimulating discussions with members of the MiniBooNE and MINER$\nu$A experiments.
The writing of this review was partially supported by Deutsche Forschungsgemeinschaft.

\appendix
\section{Recent Developments after Publication}
Since this article was written (Dec.\ 2015) some major new developments have occured that should briefly be mentioned here.
\subsection{Role of RPA Correlations}
The work of Martini et al.\ as well as that of Nieves et al.\ showed a large influence of RPA correlations which tended to lower the cross section significantly at low $Q^2$. A clear example for this effect is shown in Fig.\ 3 in Ref.\ \cite{Nieves:2011yp} where it is seen that the RPA effects essentially just neutralize the extra contribution from 2p2h excitations. Results obtained in a continuum RPA (CRPA) calculation \cite{Pandey:2016jju} seem to contradict that result. For example, Fig.\ 2 in Ref.\ \cite{Pandey:2016jju} shows that the inclusive cross section obtained with RPA correlations essentially agree with those obtained from a mean-field model, in which the mean-field was obtained from a Hartree-Fock calculation and no explicit RPA correlations are present.

This seeming disagreement can be understood when one realizes that the results of Nieves et al.\ as well as those of Martini et al.\ were obtained by using an unbound groundstate without any mean-field potential. In contrast, the calculations of Pandey et al.\ \cite{Pandey:2016jju} used a Hartree-Fock mean-field potential which binds the nucleus. Obviously then such a realistic, bound ground state, in contrast to the unbound ground state of a local Fermi gas, contains some of the effects of the RPA correlations. This observation also explains why the GiBUU results, obtained also with a realistic mean-field potential without RPA, give a good description of the QE-peak in the inclusive cross section \cite{Gallmeister:2016dnq}. The same is true if the QE cross section is obtained in the scaling approach from a relativistic mean field potential without any additional RPA effects \cite{Megias:2016fjk}. In addition, it has been pointed out that the size of RPA effects depends also on the particular NN interaction used \cite{Martini:2016eec}.

\subsection{Microscopic Theory of 2p2h Processes}
GiBUU now obtains the 2p2h part of the structure function from electron scattering under the assumption that the process is purely transverse and that the axial and the vector contributions are proportional to each other \cite{Gallmeister:2016dnq}; the same approximations are used in the calculations of Martini et al. The structure function $W_1$ is obtained from a fit to electron scattering data. The agreement reached for electron scattering as well as for neutrino-induced data from different experiments and for different neutrino flavors is very good \cite{Gallmeister:2016dnq}.

Starting from an effective NN interaction the authors of Ref.\ \cite{RuizSimo:2016ikw} have calculated the 2p2h response of nuclei to incoming neutrinos, assuming a relativistic Fermi gas for the ground state. These authors did not make any of the assumptions listed in the preceding paragraph from the outset, but their results show that the dominance of the transverse channel indeed holds. When combining their results with a semi-empirical scaling description of both the one-body QE process and the inelastic excitations they also obtain a similarly good agreement with both electron and neutrino inclusive cross sections \cite{Megias:2016fjk}. The scaling functions are based on a phenomenologically modified relativistic mean-field potential. The ground state used for the calculation of the QE and inelastic parts of the reaction is thus quite different from that used for the 2p2h part.

% References
%
% Margin notes within bibliography
%\section*{LITERATURE\ CITED}

%To download the appropriate bibliography style file, please see %\url{http://www.annualreviews.org/page/authors/author-instructions/preparing/latex}.

%\\

\bibliographystyle{ar-style5}   % if natbib is available

\end{document}